\documentclass[10pt]{article}
\usepackage{amssymb,amsmath}
\usepackage{amsthm}
\usepackage{mathrsfs}
\usepackage{graphicx}
\usepackage{dcolumn}
\usepackage{bm}
\usepackage{fullpage}
\usepackage{color}
\usepackage[all]{xy} 
\begin{document}
\newtheorem{Def}{Definition}[section]
\newtheorem{Thm}{Theorem}[section]
\newtheorem{Proposition}{Proposition}[section]
\newtheorem{Lemma}{Lemma}[section]
\theoremstyle{definition}
\newtheorem*{Proof}{Proof}
\newtheorem{Example}{Example}[section]
\newtheorem{Postulate}{Postulate}[section]
\newtheorem{Corollary}{Corollary}[section]
\newtheorem{Remark}{Remark}[section]
\theoremstyle{remark}
\newcommand{\beq}{\begin{equation}}
\newcommand{\beqa}{\begin{eqnarray}}
\newcommand{\eeq}{\end{equation}}
\newcommand{\eeqa}{\end{eqnarray}}
\newcommand{\non}{\nonumber}
\newcommand{\lb}{\label}
\newcommand{\fr}[1]{(\ref{#1})}
\newcommand{\cc}{\mbox{c.c.}}
\newcommand{\nr}{\mbox{n.r.}}
\newcommand{\Id}{\mathrm{Id}}
\newcommand{\bb}{\mbox{\boldmath {$b$}}}
\newcommand{\bbe}{\mbox{\boldmath {$e$}}}
\newcommand{\bt}{\mbox{\boldmath {$t$}}}
\newcommand{\bn}{\mbox{\boldmath {$n$}}}
\newcommand{\br}{\mbox{\boldmath {$r$}}}
\newcommand{\bC}{\mbox{\boldmath {$C$}}}
\newcommand{\bp}{\mbox{\boldmath {$p$}}}
\newcommand{\bx}{\mbox{\boldmath {$x$}}}
\newcommand{\bF}{\mbox{\boldmath {$F$}}}
\newcommand{\bT}{\mbox{\boldmath {$T$}}}
\newcommand{\balpha}{\mbox{\boldmath {$\alpha$}}}
\newcommand{\bbeta}{\mbox{\boldmath {$\beta$}}}
\newcommand{\bgamma}{\mbox{\boldmath {$\gamma$}}}
\newcommand{\bomega}{\mbox{\boldmath {$\omega$}}}
\newcommand{\bsigma}{\mbox{\boldmath {$\sigma$}}}
\newcommand{\bDelta}{\mbox{\boldmath {$\Delta$}}}
\newcommand{\bPsi}{\mbox{\boldmath {$\Psi$}}}
\newcommand{\ve}{{\varepsilon}}
\newcommand{\e}{\mathrm{e}}
\newcommand{\hF}{\widehat F}
\newcommand{\hL}{\widehat L}
\newcommand{\tA}{\widetilde A}
\newcommand{\tB}{\widetilde B}
\newcommand{\tC}{\widetilde C}
\newcommand{\tL}{\widetilde L}
\newcommand{\tK}{\widetilde K}
\newcommand{\tX}{\widetilde X}
\newcommand{\tY}{\widetilde Y}
\newcommand{\tU}{\widetilde U}
\newcommand{\tZ}{\widetilde Z}
\newcommand{\talpha}{\widetilde \alpha}
\newcommand{\te}{\widetilde e}
\newcommand{\tv}{\widetilde v}
\newcommand{\ts}{\widetilde s}
\newcommand{\tx}{\widetilde x}
\newcommand{\ty}{\widetilde y}
\newcommand{\ud}{\underline{\delta}}
\newcommand{\uD}{\underline{\Delta}}
\newcommand{\chN}{\check{N}}
\newcommand{\chX}{\check{X}}
\newcommand{\cA}{{\cal A}}
\newcommand{\cB}{{\cal B}}
\newcommand{\cC}{{\cal C}}
\newcommand{\cD}{{\cal D}}
\newcommand{\cE}{{\cal E}}
\newcommand{\cF}{{\cal F}}
\newcommand{\cH}{{\cal H}}
\newcommand{\cK}{{\cal K}}
\newcommand{\cL}{{\cal L}}
\newcommand{\cM}{{\cal M}}
\newcommand{\cR}{{\cal R}}
\newcommand{\cO}{{\cal O}}
\newcommand{\cY}{{\cal Y}}
\newcommand{\cU}{{\cal U}}
\newcommand{\cV}{{\cal V}}
\newcommand{\cZ}{{\cal Z}}
\newcommand{\tcA}{\widetilde{\cal A}}
\newcommand{\DD}{{\cal D}}
\newcommand\TYPE[3]{ \underset {(#1)}{\overset{{#3}}{#2}}  }
\newcommand{\bfe}{\boldsymbol e} 
\newcommand{\bfb}{{\boldsymbol b}}
\newcommand{\bfd}{{\boldsymbol d}}
\newcommand{\bfh}{{\boldsymbol h}}
\newcommand{\bfg}{{\boldsymbol g}}
\newcommand{\bfj}{{\boldsymbol j}}
\newcommand{\bfn}{{\boldsymbol n}}
\newcommand{\bfA}{{\boldsymbol A}}
\newcommand{\bfB}{{\boldsymbol B}}
\newcommand{\bfD}{{\boldsymbol D}}
\newcommand{\bfJ}{{\boldsymbol J}}
\newcommand{\dr}{\mathrm{d}}
\newcommand{\TE}{\mathrm{TE}}
\newcommand{\TM}{\mathrm{TM}}
\newcommand{\Ai}{\mathrm{Ai}}
\newcommand{\Bi}{\mathrm{Bi}}
\newcommand{\sech}{\mathrm{sech}}
\newcommand{\A}{   \TYPE 1  {A}  {}   }
\newcommand{\Ap}{  \TYPE p  {A}  {}   }
\newcommand{\F}{   \TYPE 2  {F}  {}   }
\newcommand{\G}{   \TYPE 2  {G}  {}   }
\newcommand{\jthree}{  \TYPE 3  {j}  {}   }
\newcommand{\cS}{  \TYPE m  {\check{S}} {}   }
\newcommand{\alp}{ \TYPE p  {\alpha}   {}   }
\newcommand{\al}[1]{ \TYPE {#1}  {\alpha}   {}   }
\newcommand{\bep}{ \TYPE p  {\beta}   {}   }
\newcommand{\be}[1]{ \TYPE {#1}  {\beta}   {}   }
\newcommand{\gamq}{ \TYPE q  {\gamma}   {}   }
\newcommand{\hash}{\#}
\newcommand{\hashat}{\widehat{\#}}
\newcommand{\hashch}{\stackrel{\vee}{\#}}
\newcommand{\chd}{\stackrel{\vee}{\D}}
\newcommand\NN[1]{{\cal N}_{#1}}
\newcommand\MM[1]{{\cal M}_{#1}}
\newcommand\BAE[1]{{\begin{equation}{\begin{aligned}#1\end{aligned}}\end{equation}}}
\newcommand{\GamCLamM}[1]{{\Gamma\mathbb{C}\Lambda^{{#1}}\cal{M}}}
\newcommand{\GamLamM}[1]{{\Gamma\Lambda^{{#1}}\,\cal{M}}}
\newcommand{\GamLamU}[1]{{\Gamma\Lambda^{{#1}}\,\cal{U}}}
\newcommand{\GamLamHU}[1]{{\Gamma\Lambda^{{#1}}\,\widehat{\cal{U}}}}
\newcommand{\GamLam}[2]{{\Gamma\Lambda^{{#1}}\,{#2}}}
\newcommand{\Lam}[2]{{\bigwedge^{{#1}}\,{#2}}}
\newcommand{\GTM}{{\Gamma T\cal{M}}}
\newcommand{\GTU}{{\Gamma T\cal{U}}}
\newcommand{\GT}[1]{{\Gamma T {#1}}}
\newcommand{\T}[1]{{T {#1}}}
\newcommand{\normM}[2]{\left(  #1\, , \, #2 \right)}
\newcommand{\normU}[2]{\left\{ #1\, , \, #2 \right\}}
\newcommand{\diag}[1]{\mbox{diag}\{\, #1\,\}}
\newcommand{\GtM}[2]{\Gamma T^{#1}_{#2}{\cal M}}
\newcommand{\EM}{\mathrm{EM}} 
\newcommand{\inp}[2]{\left\langle\,  #1\, , \, #2\, \right\rangle}
\newcommand{\equp}[1]{\overset{\mathrm{#1}}{=}}
\newcommand{\wt}[1]{\widetilde{#1}}
\newcommand{\wh}[1]{\widehat{#1}}
\newcommand{\ol}[1]{\overline{#1}}
\newcommand{\ch}[1]{\check{#1}}
\newcommand{\ii}{\imath}
\newcommand{\ic}{\sqrt{-1}}
\newcommand{\mbbD}{\mathbb{D}}
\newcommand{\mbbM}{\mathbb{M}}
\newcommand{\mbbR}{\mathbb{R}}
\newcommand{\mbbH}{\mathbb{H}}
\newcommand{\mbbT}{\mathbb{T}}
\newcommand{\mbbV}{\mathbb{ V}}
\newcommand{\mbbF}{\mathbb{F}}
\newcommand{\mbbZ}{\mathbb{Z}}
\newcommand{\Leftrightup}[1]{\overset{\mathrm{#1}}{\Longleftrightarrow}}
\newcommand{\Leg}{{\mathfrak{L}}}
\title{Maxwell's equations in media as a contact Hamiltonian vector field\\  
and its information geometry\\ 
-- An approach with a bundle whose fiber is a contact manifold --}
\author{Shin-itiro Goto\\
 Department of Applied Mathematics and Physics, \\
Graduate School of Informatics, Kyoto University,\\
Yoshida-Honmachi, Sakyo-ku, Kyoto 606-8501, Japan
}
\date{\today}
\maketitle
\begin{abstract}%
It is shown that Maxwell's equations in media without source 
can be written as a contact Hamiltonian vector field restricted to  
a Legendre submanifold, where this submanifold 
is in a fiber space of a bundle and is generated by either   
electromagnetic energy functional or co-energy functional. 
Then, it turns out that Legendre duality for this system  
gives the induction oriented formulation of Maxwell's equations  
and field intensity oriented one. Also, information geometry of 
the Maxwell fields is introduced and discussed. 
\end{abstract}%

%

\section{Introduction}
Contact geometry is often referred to as an odd-dimensional twin of 
symplectic geometry and then it has been studied from purely mathematical 
viewpoints\cite{Arnold1976}. 
On the other hand, 
there are several applications in 
science and foundation of engineering. These applications include 
equilibrium thermodynamics\cite{Hermann1973,MrugalaX,Mrugala1991,Schaft2007}, 
nonequilibrium thermodynamics\cite{Bravetti-Sep-2014,Goto2015},   
statistical mechanics\cite{Mrugala1990,Jurkowski2000,Bravetti-Aug-2014}, 
fluid mechanics\cite{Ghrist2007}, 
control theory\cite{Ohsawa2015}, 
statistical theory for non-conservative system\cite{Bravetti-JPhysA2014},
electric circuits\cite{Eberard2006},  
dissipative mechanical systems\cite{Bravetti2017},    
and so on. 
In general, if geometric theories of mathematical disciplines are ascribed to
the same geometry, then it can be expected that there are links among these 
disciplines. These links may give a unified geometric 
picture of such disciplines.  
Such an example is found in contact geometry, where 
information geometry is linked to 
contact geometric thermodynamics\cite{Goto2015,Mrugala1990}.
Here information geometry is a geometrization of parametric statistics
\cite{AN}, and it has been applied to various disciplines including 
equilibrium statistical mechanics\cite{Fujiwara2009}, control 
theory\cite{Ohara1994}, and so on.
In Ref.\cite{Goto2016}, several simple electric circuit models without 
any external source have been discussed 
in terms of contact and information geometries.   
Since 
electromagnetic media can be seen as 
distributed element models of electrical circuits, one can expect that 
Maxwell's equations without source in media 
can be written as some kind of contact geometry and 
information geometry. 

There exists a history of developing geometric formulations of 
Maxwell's equations for describing electromagnetic fields. 
The most well-known one is that these equations are described 
on a $4$-dimensional pseudo Riemannian 
manifold\cite{Nakahara,Frankel,Benn-Tucker}. 
Also, it has been shown that 
Maxwell's vacuum equations are written 
as an infinite dimensional Hamiltonian systems\cite{Abraham-Marsden-Ratiu}.
Furthermore, since  
Dirac structures on manifolds are known to be 
a geometric generalization of phase spaces of 
Hamiltonian systems\cite{Courant1990,Schaft2014},  
one expects that Maxwell's equations can be 
written with an extension of a Dirac structure.  
It then has been shown that Maxwell's equations are described as 
a Hamiltonian system with respect to a 
Stokes-Dirac structure\cite{Schaft-Maschke2002}.
Note that some wave solutions to Maxwell's equations 
can be described in terms of contact geometry\cite{Dahl2004}, and 
some electric circuits are described as a Dirac 
structure\cite{Yoshimura2006,Blankenstein2005}. 


In this paper it is shown how Maxwell's equations without source in media  
are described 
in terms of a bundle whose fiber space is a contact manifold.   
To this end, an idea of such a bundle   and  
how a contact Hamiltonian vector field 
will be formulated on such a bundle are discussed.
In this formulation, an analogue of convexity of the 
electromagnetic energy functional is emphasized, and then one notices  
that the property of such functionals allows us to use convex analysis 
so that Legendre duality should be explored in this context. 
It will then be shown that this duality leads to 
an induction field oriented formulation of Maxwell's equations and 
field intensity oriented formulation.
Also, it will turn out that the existence of such convex functional leads to  
an analogue of a dually flat space introduced in information geometry. 
These formulations can shed light on 
how Legendre transform and convex functions play a role in 
electromagnetism. 

\section{Mathematical preliminaries}
In this section mathematical symbols and tools 
are fixed and these will be used in the 
following sections.  
Mathematical objects in this paper are assumed smooth and real. 
From a viewpoint of dynamical systems theory, Maxwell's equations are 
an infinite dimensional system. To formulate these equations the 
dimension of phase space should be infinite. 
To this end, one starts with the finite dimensional case. 

\subsection{Contact geometry}
The definition of contact manifold below follows the one often used in 
contact geometric equilibrium thermodynamics ( see Ref.\,\cite{MrugalaX} ).   
\begin{Def}
(Contact manifold): 
Let $\cC$ be a $(2n+1)$-dimensional manifold $(n=1,2,\ldots)$. 
If $\cC$ carries a $1$-form $\lambda$ such that
$$
\lambda\wedge\underbrace{\dr \lambda\cdots\wedge\dr \lambda}_{n}
\neq 0,
$$
then the pair $(\cC,\lambda)$ is referred to as a ($(2n+1)$-dimensional) 
contact manifold.
\end{Def}
In this paper the following coordinate system is often used.
\begin{Thm}
\label{theorem-Darboux}
(Darboux's theorem): 
Let $(\cC,\lambda)$ be a contact manifold. 
Then there exists the local coordinate 
system $(x,p,z)$ such that $\lambda=\dr z-p_{\,a}\,\dr x^{\,a}$ with $x=\{x^{\,1},\ldots, x^{\,n}\}$ and $p=\{p_{\,1},\ldots,p_{\,n}\}$.
\end{Thm}
In this paper, Einstein notation, when allowed index variables  
appear twice in a single 
term it implies that all the values of the index are summed, is used. 
Note that,  instead of $\lambda$ in Theorem \ref{theorem-Darboux}, 
another convention exists  in the literature.  
\begin{Def}
(Darboux coordinates): 
The coordinate system stated in Theorem \ref{theorem-Darboux} is referred to as 
the Darboux coordinates or canonical coordinates.
\end{Def}

The following transformation group preserves the contact structure 
$\ker(\lambda):=\{X\in\Gamma T\cC|\ii_{\,X}\lambda=0\}$ 
for a given contact manifold $(\,\cC,\lambda\,)$. 
\begin{Def}
(Contact transformation group): 
Let $(\cC,\lambda)$ be a contact manifold and $\{\Phi_{\,\varsigma}\}:\cC\to\cC$ 
a set of elements of a group. If $\Phi_{\,\varsigma}$ is such that 
$$
\Phi_{\,\varsigma}^{\,*}\lambda
=\,f_{\,\varsigma}\lambda
$$ 
where $f_{\,\varsigma}$ is some non-vanishing function 
and $\Phi_{\,\varsigma}^{\,*}\lambda$ 
is a pull-back of $\lambda$,    
then $\{\Phi_{\,\varsigma}\}$ is referred to as a contact transformation group.
\end{Def}

In the following sections the phase spaces of Maxwell's equations for media 
will be described in terms of a bundle whose fiber space is a contact manifold. 
In particular, the phase space of such equations is  
described on a Legendre submanifold of the fiber space. 
Roughly speaking these objects are appropriate for describing 
dynamical systems as partial differential equations, in addition that 
the standard contact geometry is appropriate for describing dynamical systems 
written as     
ordinary differential equations. 

To describe dynamical systems written as ordinary differential equations  
on contact manifolds, one needs the following.

\begin{Def}
(Reeb vector field): 
Let $(\cC,\lambda)$ be a contact manifold, and $\cR$ a vector field on $\cC$.
If $\cR$ satisfies 
$$
\ii_{\cR}\lambda
=1,\quad\mbox{and}\quad 
\ii_{\cR}\dr\lambda
=0,
$$ 
then $\cR$ is referred to 
as the Reeb vector field or the characteristic vector field. 
\end{Def}
Here $\ii_X$ is the interior product associated with a vector field $X$.
It is known that the Reeb vector field is uniquely determined and 
that the Darboux coordinate expression of $\cR$ is $\partial/\partial z$.

Roughly speaking, 
the following vector fields are obtained by the infinitesimal transforms 
of contact transforms  
for a given contact manifold $(\,\cC,\lambda\,)$. 
\begin{Def}
(Contact vector field): 
Let $(\cC,\lambda)$ be a contact manifold and $X$ a vector field on $\cC$. 
If $X$ satisfies $\cL_{X}\lambda=f\,\lambda$ with some function $f$, then 
$X$ is referred to as a contact vector field. 
\end{Def}
Here $\cL_{\,X}$ denotes the Lie derivative with respect to a vector field 
$X$. 
To realize a contact vector field, one can take a contact Hamiltonian vector field. 
With the Reeb vector field, one defines the following. 
\begin{Def}
(Contact Hamiltonian vector field): 
Let $(\cC,\lambda)$ be a contact manifold, $h$ a function on $\cC$, 
and $X_h$ a vector field on $\cC$. 
If $X_h$ satisfies 
\beq
\ii_{X_h}\lambda
=h,\quad\mbox{and}\quad 
\ii_{X_h}\dr\lambda
=-\,(\dr h-(\cR\,h)\lambda),
\label{conditions-for-contact-Hamiltonian-vector}
\eeq
then $X_h$ is referred to 
as the contact Hamiltonian vector field associated with $h$, 
and $h$ a contact Hamiltonian.
\end{Def}
\begin{Remark}
 Applying 
the Cartan formula $\cL_{\,X}\alpha=\dr\ii_X\alpha+\ii_X\dr\alpha$ for 
a $q$-form field $\alpha$, one has a basic property for 
a contact Hamiltonian vector field : 
$$
\cL_{X_h}\lambda
=(\cR h)\lambda.
$$
\end{Remark}
Contact Hamiltonian vector fields are to be used 
for describing various equations on contact manifolds in the following 
sections.

By straightforward calculations one can show the following.
\begin{Proposition}
(Coordinate expression of contact Hamiltonian vector field): 
The Darboux coordinate expression of 
\fr{conditions-for-contact-Hamiltonian-vector} is given as 
$$
X_h
=\dot{x}^{\,a}\frac{\partial}{\partial x^{\,a}}
+\dot{p}_{\,a}\frac{\partial}{\partial p_{\,a}}
+\dot{z}\frac{\partial}{\partial z},
$$
where 
\beq
\dot{x}^{\,a}
=-\,\frac{\partial h}{\partial p_{\,a}},\qquad 
\dot{p}_{\,a}
=\frac{\partial h}{\partial x^{\,a}}+p_{\,a}\frac{\partial h}{\partial z},
\qquad 
\dot{z}
=h-p_{\,a}\frac{\partial h}{\partial p_{\,a}}.
\label{coordinate-expression-contact-Hamiltonian-vector}
\eeq
\end{Proposition}
Let $\phi_{\,h}$ be an integral curve of $X_{\,h}$ such that 
$\phi_{\,h}:\mbbT\to\cC, (t\mapsto (x,p,z))$ is a map with some 
$\mbbT\subseteq\mbbR$. 
Then $\dot{}$ denotes the derivative with respect to $t$. 
In this case, one has dynamical systems. 
Physically this $t\in\mbbT$ is interpreted as time.
In what follows 
contact vector fields are always treated as dynamical systems and integral 
curves are focused when a contact Hamiltonian vector field is given. 
\begin{Remark}
Unlike the case of autonomous Hamiltonian vector fields, 
contact Hamiltonian vector fields need not be conserved, since 
$\cL_{\,X_{\,h}}h=(\cR h)h$ does not vanish in general.   
\end{Remark}
In applications of contact geometry, Legendre submanifolds play various roles.
The definition of Legendre submanifold is as follows.
\begin{Thm}
(Maximal dimension of integral submanifold):  
Let $(\cC,\lambda)$ be a $(2n+1)$-dimensional contact manifold. 
The maximal dimension of integral submanifold of $\lambda=0$ is $n$.
\end{Thm}
\begin{Def}
\label{definition-Legendre-submanifold}
(Legendre submanifold): Let $(\cC,\lambda)$ be a contact manifold and $\cA$ a submanifold of $\cC$. If $\cA$ is a maximal dimensional integral submanifold of 
$\lambda$, then $\cA$ is referred to as a Legendre submanifold.  
\end{Def}
It has been known that local expressions of Legendre submanifolds are 
described by some functions on contact manifolds. 

\begin{Thm}
\label{theorem-Legendre-submanifold-theorem-Arnold}
(Local expressions of Legendre submanifolds, \cite{Arnold1976}):   
Let $(\,\cC,\lambda\,)$ be a $(2n+1)$-dimensional contact manifold,  
and $(x,p,z)$ the canonical coordinates such that 
$\lambda=\dr z-p_{\,a}\,\dr x^{\,a}$  
with $x=\{\,x^{\,1},\ldots,x^{\,n}\,\}$ and $p=\{\,p_{\,1},\ldots,p_{\,n}\,\}$. 
For any partition $I\cup J$ of the set of indices $\{\,1,\ldots,n\,\}$ into 
two disjoint subsets $I$ and $J$, and for a function $\phi(x^J,p_I)$ of 
$n$ variables $p_{\,i},i\in I$, and $x^{\,j},j\in J$ the $(n+1)$ equations
\beq
x^i=-\,\frac{\partial\phi}{\partial p_i},\qquad
p_j=\frac{\partial\phi}{\partial x^j},\qquad 
z=\phi-p_i\frac{\partial\phi}{\partial p_i},
\label{Legendre-submanifold-theorem-Arnold}
\eeq
define a Legendre submanifold. Conversely, every 
Legendre submanifold of $(\,\cC,\lambda\,)$ 
in a neighborhood of any point is    
defined by these equations for at least one of the $2^n$ possible choices 
of the subset $I$.
\end{Thm}
\begin{Def}
(Legendre submanifold generated by function):  
The function $\phi$ used in 
Theorem\,\ref{theorem-Legendre-submanifold-theorem-Arnold}  
is referred to as a generating function of the Legendre submanifold. 
If a  Legendre submanifold $\cA$ is expressed 
as \fr{Legendre-submanifold-theorem-Arnold}, then $\cA$ is referred to as 
a Legendre submanifold generated by $\phi$. 
\end{Def}

The following are examples of local expressions for 
 Legendre submanifolds.  
\begin{Example}
\label{example-Arnold-Legendre-submanifold-psi}
Let $(\,\cC,\lambda\,)$ be a $(2n+1)$-dimensional contact manifold,
$(x,p,z)$ the canonical coordinates such that 
$\lambda=\dr z-p_{\,a}\,\dr x^{\,a}$  
with $x=\{\,x^{\,1},\ldots,x^{\,n}\,\}$ and $p=\{\,p_{\,1},\ldots,p_{\,n}\,\}$, 
and $\psi$ a function of $x$ only.   
Then, the Legendre submanifold $\cA_{\,\psi}$ generated by $\psi$ with 
$\Phi_{\,\cC\cA\psi}:\cA_{\,\psi}\to\cC$ being the embedding 
is such that 
\beq
\Phi_{\,\cC\cA\psi}\cA_{\,\psi}
=\left\{\ (x,p,z)\in\cC \ \bigg|\ 
p_j=\frac{\partial\psi}{\partial x^{\,j}},\ \mbox{and}\ 
z=\psi(x),\quad j\in \{\,1,\ldots,n\,\}
\ \right\}. 
\label{example-psi-Legendre-submanifold}
\eeq
The relation between this $\psi$ and $\phi$ of   
\fr{Legendre-submanifold-theorem-Arnold} is $\psi(x)=\phi(x)$ with 
$J=\{\,1,\ldots,n\,\}$.
One can verify that $\Phi_{\,\cC\cA\psi}^{\ \ \ \  *}\lambda=0$. 
\end{Example}
\begin{Example}
\label{example-Arnold-Legendre-submanifold-varphi}
Let $(\,\cC,\lambda\,)$ be a $(2n+1)$-dimensional contact manifold,
$(x,p,z)$ the canonical coordinates such that 
$\lambda=\dr z-p_{\,a}\,\dr x^{\,a}$ 
with $x=\{\,x^{\,1},\ldots,x^{\,n}\,\}$ and $p=\{\,p_{\,1},\ldots,p_{\,n}\,\}$, 
and $\varphi$ a function of $p$ only. 
Then, the Legendre submanifold $\cA_{\,\varphi}$ generated by $-\,\varphi$ 
with $\Phi_{\,\cC\cA\varphi}:\cA_{\,\varphi}\to\cC$ being 
the embedding is such that 
\beq
\Phi_{\,\cC\cA\varphi}\cA_{\,\varphi}
=\left\{\ (x,p,z)\in\cC \ \bigg|\ 
x^{\,i}=\frac{\partial\varphi}{\partial p_{\,i}},\ \mbox{and}\ 
  z=p_{\,i}\frac{\partial\varphi}{\partial p_{\,i}}-\varphi(p),\quad i\in 
\{\,1,\ldots,n\,\}
\ \right\}. 
\label{example-varphi-Legendre-submanifold}
\eeq
The relation between this $\varphi$ and $\phi$ of   
\fr{Legendre-submanifold-theorem-Arnold} is $\varphi(p)=-\,\phi(p)$ with
$I=\{\,1,\ldots,n\,\}$.
One can verify that $\Phi_{\,\cC\cA\varphi}^{\ \ \ \ *}\lambda=0$.
\end{Example}

In contact geometry the following transform is often used.    
Note that several conventions exist in the literature. 
\begin{Def}
(Total Legendre(-Fenchel) transform): 
Let $\cM$ be an $n$-dimensional manifold,
$x=\{\,x^{\,1},\ldots,x^{\,n}\,\}$ coordinates, 
and $\psi$ a function of $x$.
Then the total Legendre transform of $\psi$ with respect to $x$ 
is defined to be  
\beq
\Leg[\psi](p)
:=\sup_{x}\left[\,x^{\,a}p_{\,a}-\psi(x)\,\right],
\label{def-total-Legendre-transform}
\eeq
where $p=\{\,p_{\,1},\ldots,p_{\,n}\,\}$.
\end{Def}
\begin{Remark}
\label{standard-contact-geometry-Legendre-diffeomorphic}
If $\psi$ in 
Example\,\ref{example-Arnold-Legendre-submanifold-psi} 
is strictly convex, and $\varphi$ 
in Example \ref{example-Arnold-Legendre-submanifold-varphi}
is chosen as $\varphi(p)=\Leg[\psi](p)$, then it follows that 
$\cA_{\,\psi}$ is diffeomorphic to  $\cA_{\,\varphi}$ 
( see Ref.\,\cite{Goto2015} ).  
\end{Remark}

Introducing some symbols, 
one can have other equivalent expressions for the Legendre submanifolds  
\fr{example-psi-Legendre-submanifold} and 
\fr{example-varphi-Legendre-submanifold}.
The following definitions were introduced in Ref.\,\cite{Goto2016}, and  
it was shown that the introduced functions are tools to describe 
contact Hamiltonian vector fields concisely. They are summarized as follows.
\begin{Def}
\label{definition-adapted-function}
(Adapted functions):  
Let $(\cC,\lambda)$ be a $(2n+1)$-dimensional contact manifold, $(x,p,z)$ 
canonical coordinates such that $\lambda=\dr z-p_{\,a}\,\dr x^{\,a}$ with 
$x=\{x^{\,1},\ldots,x^{\,n}\}$ and $p=\{p_{\,1},\ldots,p_{\,n}\}$. In addition let 
$\psi$ be a 
function on $\cC$ depending on $x$ only,   
and $\varphi$ a function on $\cC$ depending on $p$ only. Then
the functions 
$\Delta_{\,0}^{\,\psi},\{\Delta_{\,1}^{\,\psi},\ldots,\Delta_{\,n}^{\,\psi}\}:\cC\to\mbbR$
and $\Delta_{\,\varphi}^{\,0},\{\Delta_{\,\varphi}^{\,1},\ldots,\Delta_{\,\varphi}^{\,n}\}:\cC\to\mbbR$ 
such that 
$$
\Delta_{\,0}^{\,\psi}(x,z)
:=\psi(x)-z,\quad 
\Delta_{\,a}^{\,\psi}(x,p)
:=\frac{\partial\psi}{\partial x^{\,a}}-p_{\,a},\qquad 
a\in\{\,1,\ldots,n\,\}, 
$$
$$
\Delta_{\,\varphi}^{\,0}(x,p,z)
:=x^{\,j}p_{\,j}-\varphi(p)-z,\quad 
\Delta_{\,\varphi}^{\,a}(x,p)
:=x^{\,a}-\frac{\partial\varphi}{\partial p_{\,a}},\qquad 
a\in\{\,1,\ldots,n\,\}, 
$$  
are referred to as adapted functions. 
\end{Def}
In adapted functions, the local expressions of Legendre submanifolds
generated by $\psi$ and those by $-\varphi$ can be written as 
follows\cite{Goto2016}. 
\begin{Proposition}
\label{proposition-Legendre-submanifold-adapted-functions}
(Local expressions of Legendre submanifold with adapted functions,\, 
\cite{Goto2016}): 
The Legendre submanifold $\cA_{\,\psi}$ generated by $\psi$ as in 
\fr{example-psi-Legendre-submanifold}
is expressed as 
$$
\Phi_{\,\cC\cA\psi}\cA_{\,\psi}
=\left\{\ (x,p,z)\in\cC \ |\ 
\Delta_{\,0}^{\,\psi}
=0\ \mbox{and}\ 
\Delta_{\,1}^{\,\psi}
=\cdots
=\Delta_{\,n}^{\,\psi}
=0
\ \right\}, 
$$
where 
$\Phi_{\,\cC\cA\psi}\cA_{\,\psi}:\cA_{\,\psi}\to\cC$ is the embedding. Similarly, 
the Legendre submanifold $\cA_{\,\varphi}$ generated by $-\varphi$ as in 
\fr{example-varphi-Legendre-submanifold}
is expressed as 
$$
\Phi_{\,\cC\cA\varphi}\cA_{\,\varphi}
=\left\{\ (x,p,z)\in\cC \ |\ 
\Delta_{\,\varphi}^{\,0}
=0\ \mbox{and}\ 
\Delta_{\,\varphi}^{\,1}
=\cdots
=\Delta_{\,\varphi}^{\,n}
=0
\ \right\}, 
$$
where $\Phi_{\,\cC\cA\varphi}\cA_{\,\varphi}:\cA_{\,\varphi}\to\cC$ is the embedding. 
\end{Proposition} 
From this proposition, a Legendre submanifold $\Phi_{\cC\cA\psi}\cA_{\,\psi}$ 
is a submanifold where the 
constraints $\Delta_{\,0}^{\,\psi}=\cdots=\Delta_{\,n}^{\,\psi}$ hold.  
Thus a vector field on $\Phi_{\cC\cA\psi}\cA_{\,\psi}$ is the one 
where relations $\Delta_{\,0}^{\,\psi}=\cdots=\Delta_{\,n}^{\,\psi}$ hold.
This kind of a vector field can be 
constructed with a restricted contact Hamiltonian vector field.
It has been shown in Ref.\cite{Goto2016} 
that contact Hamiltonian vector fields are 
also written in terms of adapted functions.

\begin{Proposition} 
\label{Mrugala-variant-psi}
(Restricted contact Hamiltonian vector field 
as the push-forward of a vector field on 
the Legendre submanifold generated by $\psi$,\, 
\cite{Goto2016}): 
Let $\{\,F_{\,\psi}^{\,1},\ldots,F_{\,\psi}^{\,n}\,\}$ 
be a set of functions of $x$ on $\cA_{\,\psi}$ 
such that they do not identically vanish,   
and $\chX_{\,\psi}^{\,0}\in T_{\,x}\,\cA_{\,\psi}, ( x\in \cA_{\,\psi})$ 
the vector field given as   
$$
\chX_{\,\psi}^{\,0}
=\dot{x}^{\,a}\frac{\partial}{\partial x^{\,a}},\quad\mbox{where}\quad 
\dot{x}^{\,a}
=F_{\,\psi}^{\,a}(x),\qquad 
(a\in\{\,1,\ldots,n\,\}).
$$
In addition, let $X_{\,\psi}^{\,0}:=(\,\Phi_{\,\cC\cA\psi}\,)_{*}\chX_{\,\psi}^{\,0} 
\in T_{\,\xi}\cA_{\,\psi}^{\,\cC}, (\,\xi\in\cA_{\,\psi}^{\,\cC}\,)$ 
be the push-forward of
$\chX_{\,\psi}^{\,0}$,   
where $\cA_{\,\psi}^{\,\cC}:=\Phi_{\,\cC\cA\psi}\cA_{\,\psi}$ with   
$\Phi_{\,\cC\cA\psi}:\cA_{\,\psi}\to \cC$ being the embedding :  
\beqa
\Phi_{\,\cC\cA\psi} &:& \cA_{\,\psi}\to 
\cA_{\,\psi}^{\,\cC},\qquad\qquad x\mapsto (\,x,p(x),z(x)\,)
\non\\
(\,\Phi_{\,\cC\cA\psi}\,)_{\,*} &:& T_{\,x}\,\cA_{\,\psi}\to  
T_{\,\xi}\,\cA_{\,\psi}^{\,\cC},\quad \chX_{\,\psi}^{\,0}\mapsto
X_{\,\psi}^{\,0}\,.
\non
\eeqa
Then it follows that   
\beq
X_{\,\psi}^{\,0}
=\dot{x}^{\,a}\frac{\partial}{\partial x^{\,a}}
+\dot{p}_{\,a}\frac{\partial}{\partial p_{\,a}}
+\dot{z}\frac{\partial}{\partial z},\quad\mbox{where}\quad 
\dot{x}^{\,a}
=F_{\,\psi}^{\,a}(x),\quad
\dot{p}_{\,a}
=\frac{\dr}{\dr t}\left(\,\frac{\partial\psi}{\partial x^{\,a}}\,\right),
\quad 
\dot{z}
=\frac{\dr \psi}{\dr t}.
\label{tangent-vector-Legendre-submanifold-psi-component}
\eeq
In addition, 
one has that $X_{\,\psi}^{\,0}=X_{\,h_{\,\psi}}|_{\,h_{\,\psi}=0}$. Here 
$X_{\,h_{\,\psi}}$ is the contact Hamiltonian vector field associated with  
\beq
h_{\,\psi}(x,p,z)
=\Delta_{\,a}(x,p) F_{\,\psi}^{\,a}(x) 
+\Gamma_{\,\psi}(\,\Delta_{\,0}(x,z)\,),
\label{tangent-vector-Legendre-submanifold-psi-Hamiltonian}
\eeq
where $\Gamma_{\,\psi}$ is a function of $\Delta_{\,0}$ such that 
$$
\Gamma_{\,\psi}(\,\Delta_{\,0}\,)
=\left\{
\begin{array}{cl}
0&\mbox{for}\quad\Delta_{\,0}=0\\
\mbox{non-zero}&\mbox{for}\quad \Delta_{\,0}\neq 0
\end{array}
\right..
$$ 
\end{Proposition}
There exists a counterpart of Proposition\,\ref{Mrugala-variant-psi} as follows.
\begin{Proposition}
\label{Mrugala-variant-varphi}
(Restricted contact Hamiltonian vector field as the push-forward of 
vector fields on the Legendre submanifold generated by $-\,\varphi$,\, 
\cite{Goto2016}): 
Let $\{\,F_{\,1}^{\,\varphi},\ldots,F_{\,n}^{\,\varphi}\,\}$ 
be a set of functions of $p$ on $\cA_{\,\varphi}$  
such that they do not identically vanish,   
and $\chX_{\,\varphi}^{\,0}\in T_{\,p}\cA_{\,\varphi}, ( p\in\cA_{\,\varphi} )$ 
given as   
$$
\chX_{\,\varphi}^{\,0}
=\dot{p}_{\,a}\frac{\partial}{\partial p_{\,a}},\quad\mbox{where}\quad 
\dot{p}_{\,a}
=F_{\,a}^{\,\varphi}(p).
$$
In addition, let $X_{\,\varphi}^{\,0}:=(\,\Phi_{\,\cC\cA\varphi}\,)_{*}\chX_{\,\varphi}^{\,0} 
\in T_{\,\xi}\cA_{\,\varphi}^{\,\cC}, (\,\xi\in\cA_{\,\varphi}^{\,\cC}\,)$ 
be the push-forward of
$\chX_{\,\varphi}^{\,0}$,   
where $\cA_{\,\varphi}^{\,\cC}:=\Phi_{\,\cC\cA\varphi}\cA_{\,\varphi}$ with   
$\Phi_{\,\cC\cA\varphi}:\cA_{\,\varphi}\to \cC$ being the embedding : 
\beqa
\Phi_{\,\cC\cA\varphi} &:& \cA_{\,\varphi}\to 
\cA_{\,\varphi}^{\,\cC},\qquad\qquad x\mapsto (\,x(p),p,z(p)\,)
\non\\
(\,\Phi_{\,\cC\cA\varphi}\,)_{\,*} &:& T_{\,p}\,\cA_{\,\varphi}\to  
T_{\,\xi}\,\cA_{\,\varphi}^{\,\cC},\quad \chX_{\,\varphi}^{\,0}\mapsto
X_{\,\varphi}^{\,0}\,.
\non
\eeqa
Then it follows that   
\beq
X_{\,\varphi}^{\,0}
=\dot{x}^{\,a}\frac{\partial}{\partial x^{\,a}}
+\dot{p}_{\,a}\frac{\partial}{\partial p_{\,a}}
+\dot{z}\frac{\partial}{\partial z},\quad\mbox{where}\quad 
\dot{x}_{\,a}
=\frac{\dr}{\dr t}\left(\,\frac{\partial\varphi}{\partial p_{\,a}}\,\right),
\quad 
\dot{p}^{\,a}
=F_{\,a}^{\,\varphi}(p),\quad
\dot{z}
=p_{\,j}F_{\,k}^{\,\varphi}\frac{\partial^2\,\varphi}{\partial p_k\partial p_j}.
\label{tangent-vector-Legendre-submanifold-varphi-component}
\eeq
In addition, one has that $X_{\,\varphi}^{\,0}=X_{\,h_{\,\varphi}}|_{\,h_{\,\varphi}=0}$. 
Here $X_{\,h_{\,\varphi}}$ is the contact Hamiltonian vector field associated with 
\beq
h_{\,\varphi}(x,p)
=\Delta^{\,a}(x,p) F_{\,a}^{\,\varphi}(p)
+\Gamma^{\,\varphi}(\,\Delta^{\,0}\,),
\label{tangent-vector-Legendre-submanifold-varphi-Hamiltonian}
\eeq
where $\Gamma^{\,\varphi}$ is a function of $\Delta^{\,0}$ such that 
$$
\Gamma^{\,\varphi}(\,\Delta^{\,0}\,)
=\left\{
\begin{array}{cl}
0&\mbox{for}\quad\Delta^{\,0}=0\\
\mbox{non-zero}&\mbox{for}\quad \Delta^{\,0}\neq 0
\end{array}
\right..
$$ 

\end{Proposition}
\subsection{Fiber bundles}
A standard covariant form of Maxwell's equations is 
formulated on a $4$-dimensional 
pseudo Riemannian manifold, where electromagnetic fields are 
expressed in terms of a form language.   
Then the $(3+1)$-decomposition with respect to an observer of  
the covariant form of Maxwell's equations 
gives equations on a $3$-dimensional Riemannian manifold.  
By contrast, a bundle is used 
in our extended contact geometric description.  
In this subsection, various definitions and various operators for 
bundles are introduced to formulate such decomposed Maxwell's equations 
as a dynamical system in terms of a bundle formalism.
The following definition of bundle and the definitions 
of related objects are used in this paper. 
More mathematically rigorous definitions can be found in 
Ref.\cite{Nakahara} and so on.   
\begin{Def}
(Bundle or fiber bundle): 
Let $\cB$ be a $d_{\cB}$-dimensional manifold 
with local coordinates $\zeta=\{\zeta_1,\ldots,\zeta_{d_{\cB}}\}$,
$\cF$ a $d_{\cF}$-dimensional manifold, 
$\cM$ a $(d_{\cB}+d_{\cF})$-dimensional manifold, $\pi:\cM\to\cB$ a projection, 
$G$ a group acting on $\cF$, and $\{U_{\,i}\}$ an open covering of $\cB$ 
with $\phi_{\,i}:U_{\,i}\times \cF\to\pi^{-1}(U_{\,i})$ such that 
$\pi\phi_{\,i}(\zeta,u)=\zeta$.  
Then the set $(\cM,\pi,\cB)$ or $(\cM,\pi,\cB,\cF,G)$ 
 is referred to as a bundle or a fiber bundle,  
$\cB$ a base space, $\cF$ a fiber space, $G$ a structure group, 
$\phi_{\,i}$ a local trivialization, 
and $\cM$ a total space. 
Furthermore, let    
$t_{\,ij}(\zeta):=\phi_{\,i,\zeta}^{-1}\circ\phi_{\,j,\zeta}$
be an element of $G$ ($ t_{\,ij}:U_{\,i}\cap U_{\,j}\to G$ ),  
where $\phi_{\,i,\zeta}(u)=\phi_{\,i}(\zeta,u)$ 
for $U_{\,i}\cap U_{\,j}\neq \emptyset$. Then $\{t_{\,ij}\}$ are referred to as 
transition functions. 
\end{Def}  
\begin{Def}
(Trivial bundle and non-trivial bundle):  
If a transition function for a bundle can be chosen to 
be identical, then
the bundle is referred to as a trivial bundle. Otherwise, the bundle is 
referred to as a non-trivial bundle. 
\end{Def}

Non-identical transition functions are used 
for describing non-trivial bundles.   
For example, 
the M\"obius band can be constructed with this formulation\cite{Nakahara}.   
In this paper trivial bundles are only considered.

A special class of sub-space of a bundle is considered in this paper, and 
the definition is given as follows.
\begin{Def}
(Sub-bundle):
Let $(\cM,\pi,\cB)$ and $(\cM',\pi',\cB)$ be bundles. 
If the two conditions, 
\begin{enumerate}
\item
$\cM'$ is a submanifold of $\cM$, 
\item
$\pi'=\pi|_{\cM'}$
\end{enumerate}
are satisfied, then $(\cM',\pi',\cB)$ is referred to as a sub-bundle of 
$(\cM,\pi,\cB)$.
\end{Def}

\begin{Def}
(Section): 
Let $(\cM,\pi,\cB)$ be a bundle, and $f:\cB\to\cM$ a map  
such that $\pi\circ f=\Id_{\,\cB}$. Then  
 $f$ is referred to as a section. 
The space of sections is denoted\, $\Gamma\cM$. If $\cM=\Omega^{\,0}\cB$, then 
$f\in\Gamma\Lambda^{\,0}\cB$. Similarly 
the space of $q$-forms on $\cB$ is denoted  
$\Gamma\Lambda^{\,q}\cB$, and the space of vector fields on $\cB$ as 
$\Gamma T\cB$.
\end{Def}

Given a bundle $(\cM,\pi,\cB)$, 
the space of $q$-forms on $\cB$ and 
the space of $q'$-forms on $\cM$ can be introduced as follows.
\begin{Def}
(Horizontal forms): 
Let $(\cM,\pi,\cB)$ be a bundle with $\dim\cB=d_{\cB}$ and 
$\dim\cM=d_{\cB}+d_{\cF}$,  
$\zeta$ coordinates for $\cB$ with 
$\zeta=\{\zeta^{\,1},\ldots,\zeta^{\,d_{\cB}}\}$, 
$(\zeta,u)$ coordinates for $\cM$ with $u=\{\,u^{\,\,1},\ldots,u^{\,d_{\cF}}\,\}$, 
and $\{\alpha_{i_{\,1}\cdots i_{\,q}}\}\in\Gamma\Lambda^{0}\cM$ some functions. 
A $q$-form on the bundle $(\cM,\pi,\cB)$ of the form   
$$
\alpha_{\mbbH}
=\alpha_{i_{\,1}\cdots i_{\,q}}(\zeta,u)\,
\dr \zeta^{\,i_1}\wedge\cdots\wedge \dr\zeta^{i_q},
$$
is referred to as a horizontal $q$-form.
The space of horizontal $q$-forms is 
denoted $\Gamma\Lambda_{\,\mbbH}^{\,q}\cM$.
\end{Def}  
\begin{Def}
(Vertical forms): 
Let $(\cM,\pi,\cB)$ be a bundle with   
$\dim\cB=d_{\cB}$ and 
$\dim\cM=d_{\cB}+d_{\cF}$,  
$\zeta$ a set of coordinates for $\cB$, 
$(\zeta,u)$ a set of coordinates for $\cM$ 
with $\zeta=\{\zeta^{\,1},\ldots,\zeta^{\,d_{\,\cB}}\}$ and 
$u=\{u^{\,1},\ldots,u^{\,d_{\,{\cF}}}\}$, and  
$\{\alpha_{\,i_1\cdots i_q}\}\in\GamLam{0}{\cM}$ some functions.
A $q$-form on the bundle $(\cM,\pi,\cB)$ 
of the form   
$$
\alpha_{\,\mbbV}
=\alpha_{\,i_1\cdots i_q}(\zeta,u)\,
\dr u^{\,i_1}\wedge\cdots\wedge \dr u^{\,i_q},
$$
is referred to as a vertical $q$-form. The space of vertical $q$-forms is 
denoted  $\Gamma\Lambda_{\,\mbbV}^{\,q}\cM$. In addition, vertical $0$-forms are  
referred to as vertical functions.   
\end{Def}  

The wedge product of a horizontal $q$-form and a vertical $q'$-form 
can be defined. Then one defines the following. 
\begin{Def}
(Mixed form):
If a $(q+q')$-form $\alpha_{\,\mbbM}\in\Gamma\Lambda^{q+q'}\cM$ can be written as 
$$
\alpha_{\,\mbbM}
=\beta_{\,\mbbH}\wedge\gamma_{\,\mbbV},
$$
with some $\beta_{\,\mbbH}\in\Gamma\Lambda_{\,\mbbH}^{\,q}\cM$ and 
$\gamma_{\,\mbbV}\in\Gamma\Lambda_{\,\mbbV}^{\,q'}\cM$, then 
$\alpha_{\,\mbbM}$ 
is referred to as a mixed $(q,q')$-form. 
The space of mixed $(q,q')$-forms on $\cM$ is denoted  
$\Gamma\Lambda_{\,\mbbH,\mbbV}^{\,q,q'}\cM$.
\end{Def}
\begin{Def}
(Vertical derivative):
Let $\alpha_{\,\mbbV}\in\Gamma\Lambda_{\,\mbbV}^{\,q}\cM$ be 
a vertical $q$-form whose 
local expression is 
$$
\alpha_{\,\mbbV}
=\alpha_{\,i_1\cdots i_q}(\zeta,u)\,\dr u^{\,i_1}\wedge\cdots\wedge \dr u^{\,i_q}.
$$
The operator  
$\dr_{\,\mbbV}:\Gamma\Lambda_{\,\mbbV}^{q}\cM\to \Gamma\Lambda_{\,\mbbV}^{\,q+1}\cM$
whose action is such that  
$$
\dr_{\,\mbbV}\alpha_{\,\mbbV}
=\frac{\partial\alpha_{\,i_1\cdots i_q}(\,\zeta,u)}{\partial u^{\,i_0}}\dr u^{\,i_0}
\wedge
\dr u^{\,i_1}\wedge\cdots\wedge \dr u^{\,i_q},
$$
is referred to as the vertical derivative or the vertical exterior derivative. 
\end{Def}
\begin{Def}
(Functional):
Let $(\cM,\pi,\cB)$ be a bundle, 
$\alpha_{\,\mbbH}\in\Gamma\Lambda_{\,\mbbH}^{\,q}\cM$ a horizontal 
$q$-form, 
$h\in\Gamma\Lambda_{\,\mbbV}^{\,0}\cM$ a vertical $0$-form,  and 
$\cB_{\,0}\subseteq\cB$ a $q$-dimensional space. 
The integral over $\cB_{\,0}$ 
$$
\wt{h}_{\,\cB_{0}}
=\int_{\cB_{\,0}}h\,\alpha_{\,\mbbH},
$$
is referred to as a functional. The space of 
functionals  is denoted as $\Gamma F\cM$.
\end{Def}

The functional derivative 
has been used in the infinite dimensional Hamiltonian 
formulation of Maxwell's equations\cite{Abraham-Marsden-Ratiu}. 
In this paper this derivative can  also be used 
to describe Maxwell's equations as a dynamical system.
\begin{Def}
\label{definition-functional-derivative}
(Functional derivative): 
Let $(\cM,\pi,\cB)$ be a bundle, $\alpha_{\,\mbbM}$ a mixed $(q,0)$-form on 
a $q^{\,\prime}$-dimensional submanifold of $\cB$, $\cB_{\,0}\subseteq\cB$ a 
$q^{\,\prime}$-dimensional subspace, 
$\wt{h}_{\,\cB_{\,0}}$ a functional depending on $\alpha_{\,\mbbM}$, 
and $\eta\in\mathbb{R}$ a constant.  
Then the mixed 
$(q^{\,\prime}-q,0)$-form 
$\delta \wt{h}_{\,\cB_{\,0}}/\delta\alpha_{\,\mbbM}
\in\Gamma\Lambda_{\,\mbbH,\mbbV}^{\,q^{\,\prime}-q,0}\cM$ 
that is uniquely obtained by 
$$
\wt{h}_{\,\cB_{\,0}}
\left[\,\alpha_{\,\mbbM}+\eta\,\alpha_{\,\mbbM}^{\prime}\,\right]
=\wt{h}_{\,\cB_{\,0}}\left[\,\alpha_{\,\mbbM}\,\right]
+\eta\int_{\cB_{\,0}}\frac{\delta \wt{h}_{\,\cB_{\,0}}}{\delta\alpha_{\,\mbbM}}
\wedge\alpha_{\,\mbbM}^{\prime}
+\cO(\,\eta^2\,),\qquad \forall\, \alpha_{\,\mbbM}^{\prime}
\in\Gamma\Lambda_{\,\mbbH,\mbbV}^{\,q,0}\,\cM
$$
is referred to as 
the functional derivative of $\wt{h}_{\,\cB_{\,0}}$ 
with respect to $\alpha_{\,\mbbM}$. 
\end{Def}

Similar to the case of forms, 
the space of vector fields on $\cB$ and 
the space of vector fields on $\cM$ can be introduced as follows.

\begin{Def}
(Horizontal vector field): 
Let $(\cM,\pi,\cB)$ be a bundle   
with $\dim\cB=d_{\cB}$ and $\dim\cM=d_{\cB}+d_{\cF}$,  
$\zeta$ a set of coordinates for $\cB$ with 
$\zeta=\{\zeta^{\,1},\ldots,\zeta^{\,d_{\cB}}\}$, 
$(\zeta,u)$ a set of coordinates for $\cM$ 
with $u=\{u^{\,1},\ldots,u^{\,d_{\cF}}\}$, and 
$\{Y_{\,1},\ldots,Y_{\,d_{\cB}}\}\in\GamLam{0}{\cM}$ some functions.  
A vector field on the bundle $(\cM,\pi,\cB)$ of the form   
$$
Y_{\mbbH}
=Y_{i}(\zeta,u) \frac{\partial}{\partial\zeta^{\,i}},
$$
is referred to as a horizontal vector field.
The space of horizontal vector fields is 
denoted as $\Gamma T_{\,\mbbH}\cM$.
\end{Def}  
\begin{Remark}
The dual of a horizontal vector field is a horizontal $1$-form.
\end{Remark}

\begin{Def}
(Vertical vector field): 
Let $(\cM,\pi,\cB)$ be a bundle   
with $\dim\cB=d_{\,\cB}$ and $\dim\cM=d_{\,\cB}+d_{\,\cF}$,  
$\zeta$ a set of coordinates for $\cB$ with 
$\zeta=\{\,\zeta^{\,1},\ldots,\zeta^{\,d_{\cB}}\,\}$, 
$(\zeta,u)$ a set of coordinates for $\cM$ with 
$u=\{\,u^{\,\,1},\ldots,u^{\,d_{\,\cF}}\,\}$, and 
$\{Y_{\,1},\ldots,Y_{\,d_{\cF}}\}\in\Gamma\Lambda^{\,0}\cM$ some functions.  
A vector field on the bundle $(\cM,\pi,\cB)$ of the form   
$$
Y_{\,\mbbV}
=Y_{\,i}(\zeta,u) \frac{\partial}{\partial u^{\,i}},
$$
is referred to as a vertical vector field.
The space of vertical vector fields is 
denoted as $\Gamma T_{\,\mbbV}\cM$.
\end{Def}  
\begin{Remark}
The dual of a vertical vector field is a vertical $1$-form.
\end{Remark}

The action of the interior product with respect to a vertical 
vector field $Y_{\mbbV}$ for a vertical $q$-form $\alpha_{\mbbV}$  
is denoted $\ii_{Y_{\mbbV}}\alpha_{\mbbV}$ and is similar to the action of a vector 
field $Y$ to a $q$-form $\alpha$, $\ii_Y\alpha$.  

 \begin{Def}
(Interior product associated with a vertical vector field for vertical form): 
Let $(\cM,\pi,\cB)$ be a bundle with $\dim\cB=d_{\cB}$ 
and $\dim\cM=d_{\cB}+d_{\cF}$, $\zeta$ a set of coordinates for 
$\cB$ with $\zeta=\{\zeta^{\,1},\ldots,\zeta^{\,d_{\cB}}\}$, 
$\beta_{\,\mbbH}\in\Gamma\Lambda_{\,\mbbH}^{\,q}\cM$ a horizontal $q$-form,
$\gamma_{\,\mbbV}\in\Gamma\Lambda_{\,\mbbV}^{\,q'}\cM$ a vertical $q'$-form,
$Y_{\,\mbbV}\in\Gamma T_{\,\mbbV}\cM$ a vertical vector field, and 
$\alpha_{\,\mbbM}\in\Gamma\Lambda_{\,\mbbH,\mbbV}^{\,q,q'}\cM$ a mixed $(q,q')$-form 
 written as 
$$
\alpha_{\,\mbbM}
=\gamma_{\,\mbbV}\wedge\beta_{\,\mbbH}.
$$
Then the  action of $\ii_{\,Y_{\mbbV}}$ to $\alpha_{\,\mbbM}$, 
$\ii_{\,Y_{\,\mbbV}}:\Gamma\Lambda_{\mbbH,\mbbV}^{q,q'}\cM\to \Gamma\Lambda_{\mbbH,\mbbV}^{q,q'-1}\cM$
 is defined as 
$$
\ii_{\,Y_{\mbbV}}\alpha_{\,\mbbM}
=(\,\ii_{\,Y_{\mbbV}}\gamma_{\,\mbbV}\,)\wedge\beta_{\,\mbbH}.
$$
 \end{Def}

\section{Contact manifold over base space }

In this paper a contact manifold over a base space is treated as a bundle. 
\begin{Def}
(Contact manifold over a base space): 
Let $\cB$ be a $d_{\cB}$-dimensional  manifold, 
$(\cK,\pi,\cB)$ a bundle over the base space $\cB$,
the fiber space $\pi^{-1}(\zeta)$ at a point $\zeta$ of $\cB$ 
a $(2n+1)$-dimensional manifold $\cC_{\zeta}$, 
$\cK=\bigcup_{\zeta\in\cB}\cC_{\zeta}$, and 
the structure group $G$ a contact transformation group.
If $\cK$ carries a vertical form $\lambda_{\,\mbbV}$ 
such that
$$
\lambda_{\,\mbbV}\wedge\underbrace{\dr_{\,\mbbV} \lambda_{\,\mbbV}\wedge
\cdots\wedge\dr_{\,\mbbV}
\lambda_{\,\mbbV}}_{n}
\neq 0,\qquad\mbox{at each point of $\pi^{-1}(\zeta)$ at each point $\zeta$ of 
$\cB$} 
$$
then 
$\cC_{\,\zeta}$ is referred to as a ($(2n+1)$-dimensional) contact manifold 
on the fiber space $\pi^{-1}(\zeta),(\zeta\in\cB)$, 
the quadruplet $(\cK,\lambda_{\,\mbbV},\pi,\cB)$ 
is referred to as a ($(2n+1)$-dimensional) 
contact manifold over the base space $\cB$, 
and $\lambda_{\,\mbbV}$ a contact vertical form.
\end{Def}

In this paper we only consider trivial bundles,   
then the transition function is identical since this simple case is  
enough for our contact geometric formulation of 
Maxwell's equations without source in media.  
The contact geometry of the vertical space 
is the same as the standard contact geometry.    
Thus, all of the definitions and theorems for the standard contact geometry 
can be brought to vertical spaces. 
They are shown below.

At each base point $\zeta$ of $\cB$, 
one has Darboux's theorem for $\pi^{-1}(\zeta)$. Therefore one has the 
following.
\begin{Thm}
\label{theorem-Darboux-bundle}
(Existence of Darboux coordinates on fiber space):
For a $(2n+1)$-dimensional contact manifold over a base space 
$(\cK,\lambda_{\,\mbbV},\pi,\cB)$, 
there exist local coordinates $(x,p,z)$ for $\pi^{-1}(\zeta)$ 
with $x=\{\,x^{\,1},\ldots,x^{\,n}\,\}$ and 
$p=\{\,p_{\,1},\ldots,p_{\,n}\,\}$ 
in which $\lambda_{\,\mbbM}^{\,q}\in\Gamma\Lambda_{\mbbH,\mbbV}^{q,1}\cK$ 
has the form
$$
\lambda_{\,\mbbM}^{\,q}
=\rho^{\,q}\wedge\lambda_{\,\mbbV},\qquad\mbox{where}\qquad
\lambda_{\,\mbbV}
=\dr_{\,\mbbV}z-p_{\,a}\dr_{\,\mbbV}x^{\,a},
$$  
with some $\rho^{\,q}\in\Gamma\Lambda_{\,\mbbH}^{\,q}\cK$ being nowhere vanishing.
\end{Thm}
\begin{Def}
(Canonical coordinates or Darboux coordinates): 
The $(2n+1)$ coordinates introduced in Theorem\,\ref{theorem-Darboux-bundle}
are referred to as the 
canonical coordinates for a fiber space 
or the Darboux coordinates for a fiber space.  
\end{Def}

\begin{Def}
(Canonical contact mixed form and canonical contact vertical form):
Let $(\cK,\lambda_{\,\mbbV},\pi,\cB)$  be a contact manifold over  
a base space $\cB$, and $\rho^{\,q}$ a nowhere vanishing horizontal $q$-form. 
A mixed $(q,1)$-form $\lambda_{\,\mbbM}^{\,q}\in\Gamma\Lambda_{\,\mbbH,\mbbV}^{\,q,1}\cK$ 
written as 
$$
\lambda_{\,\mbbM}^{\,q}
=\rho^{\,q}\wedge\lambda_{\,\mbbV}\,,
\qquad\mbox{where}\quad
\lambda_{\,\mbbV}
=\dr_{\,\mbbV} z-p_{\,a}\dr_{\,\mbbV} x^{\,a},
$$
is referred to as the canonical contact mixed $(q,1)$-form associated with 
$\rho^{\,q}$, and $\lambda_{\,\mbbV}\in\Gamma\Lambda_{\,\mbbV}^{\,1}\cK$ 
the canonical contact vertical form.
\end{Def}
\begin{Remark}
With the canonical contact vertical form 
$\lambda_{\,\mbbV}\in\Gamma\Lambda_{\,\mbbV}^{\,1}\cK$ and a nowhere vanishing 
horizontal $d_{\,\cB}$-form $\rho^{\,d_{\,\cB}}$,   
the mixed $(d_{\,\cB},2n+1)$-form 
$$
\rho^{\,d_{\,\cB}}\wedge\lambda_{\,\mbbV}\wedge 
\underbrace{\dr_{\,\mbbV}\lambda_{\,\mbbV}\wedge\cdots\wedge
\dr_{\,\mbbV}\,\lambda_{\,\mbbV}}_{n} 
\qquad\in\Gamma\Lambda_{\,\mbbH,\mbbV}^{\,d_{\,\cB},2n+1}\cK
$$
is a volume-form on $\cK$.
\end{Remark}

\begin{Def}
(Reeb vertical vector field): 
Let $(\cK,\lambda_{\,\mbbV},\pi,\cB)$
be a contact manifold over a base space $\cB$, 
and $\cR_{\,\mbbV}$ a vertical vector field on $\cK$. If 
$\cR_{\,\mbbV}$ satisfies
$$
\ii_{\,\cR_{\,\mbbV}}\lambda_{\,\mbbV}
=1,\qquad\mbox{and}\quad
\ii_{\,\cR_{\,\mbbV}}\,\dr_{\,\mbbV}\lambda_{\,\mbbV}
=0,
$$
then $\cR_{\,\mbbV}$ is referred to as the Reeb 
vertical vector field on $\cK$. 
\end{Def}
\begin{Proposition}
(Coordinate expression of the Reeb vertical vector field): 
Let 
$(\cK,\lambda_{\,\mbbV},\pi,\cB)$
be a contact manifold over a base space $\cB$, 
and $\cR_{\,\mbbV}$ the Reeb vertical vector field on $\cK$, and 
$(x,p,z)$ the canonical coordinates such that 
$\lambda_{\,\mbbV}=\dr_{\,\mbbV}z-p_{\,a}\dr_{\,\mbbV}x^{\,a}$. 
Then the coordinate expression of the Reeb vector field is 
$$
\cR_{\,\mbbV}
=\frac{\partial}{\partial z}.
$$
\end{Proposition}

\begin{Def}
(Contact Hamiltonian vertical vector field): 
Let 
$(\cK,\lambda_{\mbbV},\pi,\cB)$
be a contact manifold over a base space $\cB$ with $\dim\cB=d_{\cB}$,  
$\cB_{\,0}\subseteq\cB$ a $d_{\cB}$-dimensional space,
$\rho^{\,d_{\cB}}\in\Gamma\Lambda_{\mbbH}^{d_{\cB}}\cK$ a nowhere vanishing 
horizontal form, 
$\cR_{\mbbV}$ the Reeb vertical vector field on $\cK$,  
$\wt{h}\in\Gamma F\cK$ the functional given by   
$$
\wt{h}
=\int_{\cB_{\,0}}h\,\rho^{\,d_{\cB}},
$$  
with some $h\in\Gamma\Lambda_{\,\mbbV}^{\,0}\cK$, and   
$X_{\,\wt{h}}$ a vertical vector field on $\cK$. 
If $X_{\,\wt{h}}$ satisfies 
\beq
\ii_{\,X_{\,\wt{h}}}\lambda_{\mbbV}
=h
\quad\mbox{and}\quad 
\ii_{\,X_{\,\wt{h}}}\,\dr_{\,\mbbV}\lambda_{\,\mbbV}
=-\,(\,\dr_{\,\mbbV}\, h-(\cR_{\,\mbbV}\,h\,)\,
\lambda_{\,\mbbV}\,),
\label{conditions-for-contact-Hamiltonian-vector-bundle}
\eeq
then $X_{\,\wt{h}}$ is referred to as 
the contact Hamiltonian vertical vector field, 
$\wt{h}$ a contact Hamiltonian functional, and 
$h$ a contact Hamiltonian vertical function.
\end{Def}
\begin{Remark}
With the Cartan formula, one has that 
$\cL_{\,X_{\,\wt{h}}}\lambda_{\,\mbbV}=(\cR_{\,\mbbV}h)\lambda_{\,\mbbV}$. Thus,   
$\cL_{\,X_{\,\wt{h}}}\lambda_{\,\mbbM}^{\,d_{\cB}}=\rho^{\,d_{\,\cB}}\wedge\cL_{\,X_{\,\wt{h}}}\lambda_{\,\mbbV}
=(\cR_{\,\mbbV}h)\lambda_{\,\mbbM}^{\,d_{\cB}}$.  
\end{Remark}

In the following the coordinate expression of 
a contact Hamiltonian vertical vector field is shown.
\begin{Proposition}
(Coordinate expression of a contact vertical Hamiltonian vector field):
Let 
$(\cK,\lambda_{\,\mbbV},\pi,\cB)$
be a contact manifold over a base space $\cB$ with $\dim\cB=d_{\,\cB}$,  
$\cB_{\,0}\subseteq\cB$ a $d_{\,\cB}$-dimensional space, 
$\rho^{\,d_{\,\cB}}\in\Gamma\Lambda_{\mbbH}^{\,d_{\cB}}\cK$ 
a nowhere vanishing form, $(x,p,z)$ 
the canonical coordinates for the fiber space such that 
$\lambda_{\,\mbbV}=\dr_{\,\mbbV}z-p_{\,a}\dr_{\,\mbbV}x^{\,a}$ with 
$x=\{\,x^{\,1},\ldots,x^{\,n}\,\}$ and $p=\{\,p_{\,1},\ldots,p_{\,n}\,\}$, 
$\wt{h}$ the contact Hamiltonian functional given by 
$$
\wt{h}
=\int_{\cB_{\,0}}h\,\rho^{\,d_{\,\cB}},
$$    
with some $h\in\Gamma\Lambda_{\,\mbbV}^{\,0}\cK$ depending on $(x,p,z)$, and
$X_{\,\wt{h}}$ the contact Hamiltonian vertical vector field on $\cK$. 
  
Then, the canonical coordinate expression of 
\fr{conditions-for-contact-Hamiltonian-vector-bundle} is given as 
$$
X_{\,\wt{h}}
=\dot{x}^{\,a}\frac{\partial}{\partial x^{\,a}}
+\dot{p}_{\,a}\frac{\partial}{\partial p_{\,a}}
+\dot{z}\frac{\partial}{\partial z},
$$
where 
\beq
\dot{x}^{\,a}
=-\,\frac{\partial h }{\partial p_{\,a}},\qquad 
\dot{p}_{\,a}
=\frac{\partial h }{\partial x^{\,a}}
+p_{\,a}\frac{\partial h}{\partial z},
\qquad 
\dot{z}
=h-p_{\,a}\frac{\partial h}{\partial p_{\,a}},
\label{coordinate-expression-contact-Hamiltonian-vertical-vector}
\eeq
or equivalently, 
$$
\dot{x}^{\,a}
=-\,\frac{\delta \wt{h} }{\delta p_{\,a}},\qquad 
\dot{p}_{\,a}
=\frac{\delta \wt{h} }{\delta x^{\,a}}
+p_{\,a}\frac{\delta\wt{h} }{\delta z},
\qquad 
\dot{z}
=h-p_{\,a}\frac{\delta\wt{h}}{\delta p_{\,a}}.
$$
\end{Proposition}
\begin{Remark}
The coordinate expression 
\fr{coordinate-expression-contact-Hamiltonian-vertical-vector} 
is formally the same as that of 
\fr{coordinate-expression-contact-Hamiltonian-vector}.
\end{Remark}

Analogous to 
Definition\,\ref{definition-Legendre-submanifold}, 
Legendre submanifold on a  bundle is defined as follows.  
\begin{Def}
(Legendre submanifold of vertical space and that of fiber space) :  
Let $(\cK,\lambda_{\,\mbbV},\pi,\cB)$ 
be a contact manifold over a base space $\cB$. 
If $\cA_{\,\zeta}$ is a maximal dimensional integral submanifold of 
$\lambda_{\,\mbbV}$ on $\pi^{-1}(\zeta)$,  $(\zeta\in\cB)$, 
then $\cA_{\,\zeta}$ 
is referred to as a Legendre submanifold
in the fiber space $\pi^{-1}(\zeta)$, and 
$\cA^{\,\cK}=\bigcup_{\,\zeta\in\cB}\cA_{\,\zeta}$ 
a Legendre submanifold of the fiber space.
\end{Def}

An analogous theorem from 
Theorem\,\ref{theorem-Legendre-submanifold-theorem-Arnold}
holds for our bundles. 
Then 
examples of Legendre submanifolds on fiber spaces are as follows.  
\begin{Example}
Let $(\,\cK,\lambda_{\,\mbbV},\pi,\cB\,)$ 
be a $(2n+1)$-dimensional contact manifold over a base space $\cB$ 
with $\dim\cB=d_{\,\cB}$, 
$(x,p,z)$ the canonical coordinates for the fiber space such that 
$\lambda_{\,\mbbV}=\dr_{\,\mbbV} z-p_{\,a}\,\dr_{\mbbV} x^{\,a}$  
with $x=\{\,x^{\,1},\ldots,x^{\,n}\,\}$ and $p=\{\,p_{\,1},\ldots,p_{\,n}\,\}$, 
$\psi\in\Gamma\Lambda_{\,\mbbV}^{\,0}\cK$ a vertical function of $x$, 
$\rho^{\,d_{\,\cB}}$ a nowhere vanishing horizontal $d_{\,\cB}$-form, 
$\cB_{\,0}\subseteq\cB$ a $d_{\cB}$-dimensional space, 
and $\wt{\psi}_{\,\cB_{\,0}}\in\Gamma F\cK$ 
the functional 
$$
\wt{\psi}_{\,\cB_0}
=\int_{\,\cB_0}\psi\,\rho^{\,d_{\,\cB}}.
$$   
Then, the Legendre submanifold $\cA_{\,\zeta\psi}$ generated 
by $\psi$ in $\pi^{-1}(\zeta),(\zeta\in\cB)$ with 
$\Phi_{\,\cC_{\zeta}\cA_{\,\zeta\psi}}:\cA_{\,\zeta\psi}\to\cC_{\zeta}$ 
being the embedding is such that 
\beq
\Phi_{\,\cC_{\zeta}\cA_{\,\zeta\psi}}\cA_{\,\zeta\psi}
=\left\{\ (x,p,z)\in\cC_{\zeta} \ \bigg|\ 
p_j=\frac{\partial\psi}{\partial x^{\,j}},\ \mbox{and}\ 
z=\psi(x),\quad j\in \{\,1,\ldots,n\,\}
\ \right\}. 
\label{example-psi-Legendre-submanifold-bundle}
\eeq
This can also be written as 
$$
\Phi_{\,\cC_{\zeta}\cA_{\,\zeta\psi}}\cA_{\,\zeta\psi}
=\left\{\ (x,p,z)\in\cC_{\zeta} \ \bigg|\ 
p_j=\frac{\delta\wt{\psi}_{\,\cB_0}}{\delta x^{\,j}},\ \mbox{and}\ 
z=\psi(x),\quad j\in \{\,1,\ldots,n\,\}
\ \right\}. 
$$
In addition, 
$(\cA_{\,\psi}^{\,\cK},\pi|_{\,\cA_{\,\psi}^{\,\cK}},\cB)$ is a sub-bundle of 
$(\cK,\pi,\cB)$, where
$$
\cA_{\,\psi}^{\,\cK}
=\bigcup_{\zeta\in\cB}\Phi_{\,\cC_{\zeta}\cA_{\,\zeta\psi}}\cA_{\,\zeta\psi}.
$$

\end{Example}

\begin{Example}
Let $(\,\cK,\lambda_{\mbbV},\pi,\cB\,)$ 
be a $(2n+1)$-dimensional contact manifold over a base space $\cB$, 
$(x,p,z)$ the canonical coordinates for the fiber space such that 
$\lambda_{\,\mbbV}=\dr_{\,\mbbV} z-p_{\,a}\,\dr_{\,\mbbV} x^{\,a}$  
with $x=\{\,x^{\,1},\ldots,x^{\,n}\,\}$ and $p=\{\,p_{\,1},\ldots,p_{\,n}\,\}$, 
$\varphi\in\Gamma\Lambda_{\,\mbbV}^{0}\cK$ a vertical function of $p$,  
$\rho^{\,d_{\,\cB}}$ a nowhere vanishing horizontal $d_{\,\cB}$-form, 
$\cB_{\,0}\subseteq\cB$ a $d_{\,\cB}$-dimensional space, 
and $\wt{\varphi}_{\,\cB_{\,0}}\in\Gamma F\cK$ 
the functional  
$$
\wt{\varphi}_{\,\cB_{\,0}}
=\int_{\,\cB_0}\varphi\,\rho^{\,d_{\cB}}.
$$   
Then, the  Legendre submanifold $\cA_{\,\zeta\varphi}$ generated 
by $-\varphi$ in $\pi^{-1}(\zeta),(\zeta\in\cB)$ with 
$\Phi_{\,\cC_{\zeta}\cA_{\,\zeta\varphi}}:\cA_{\,\zeta\varphi}\to\cC_{\,\zeta}$ 
being the embedding is such that 
\beq
\Phi_{\,\cC_{\,\zeta}\cA_{\,\zeta}\varphi}\cA_{\,\zeta\varphi}
=\left\{\ (x,p,z)\in\cC_{\zeta} \ \bigg|\ 
x_{\,i}=\frac{\partial\varphi}{\partial p_{\,i}},\ \mbox{and}\ 
z=p_{\,i}\frac{\partial\varphi}{\partial p_{\,i}}-\varphi(p),
\quad i\in \{\,1,\ldots,n\,\}
\ \right\}. 
\label{example-varphi-Legendre-submanifold-bundle}
\eeq
This can also be written as 
$$
\Phi_{\,\cC_{\,\zeta}\cA_{\,\zeta\varphi}}\cA_{\,\zeta\varphi}
=\left\{\ (x,p,z)\in\cC_{\zeta} \ \bigg|\ 
x_{\,i}=\frac{\delta\wt{\varphi}_{\,\cB_{\,0}}}{\delta p_{\,i}},\ \mbox{and}\ 
z=p_{\,i}\frac{\delta\wt{\varphi}_{\,\cB_{\,0}}}{\delta p_i}-\varphi(p),
\quad i\in \{\,1,\ldots,n\,\}
\ \right\}. 
$$
In addition, 
$(\cA_{\,\varphi}^{\,\cK},\pi|_{\,\cA_{\,\varphi}^{\,\cK}},\cB)$ is a sub-bundle of 
$(\cK,\pi,\cB)$, where 
$$
\cA_{\,\varphi}^{\,\cK}
=\bigcup_{\,\zeta\in\cB}\Phi_{\,\cC_{\,\zeta}\cA_{\,\zeta\varphi}}\cA_{\,\zeta\varphi}.
$$

\end{Example}

Although the following could not be commonly used in the literature,  
the total Legendre transform of a 
functional is defined as follows  in this paper. 
\begin{Def}
\label{definition-Legendre-transform-functional}
(Total Legendre transform of functional): 
Let 
$(\cK,\lambda_{\,\mbbV},\pi,\cB)$
be a contact manifold over a base space $\cB$ with $\dim\cB=d_{\,\cB}$,  
$\cB_{\,0}\subseteq\cB$ a $d_{\,\cB}$-dimensional space,
$\rho^{\,d_{\,\cB}}\in\Gamma\Lambda_{\,\mbbH}^{\,d_{\,\cB}}\cK$ 
a nowhere vanishing horizontal 
form, $\psi\in\Gamma\Lambda_{\,\mbbV}^{\,0}\cK$ a vertical $0$-form, 
$\wt{\psi}_{\,\cZ_{\,0}}$ a functional such that  
$$
\wt{\psi}_{\,\cZ_{\,0}}
=\int_{\cZ_{\,0}}\,\psi\,\rho^{\,d_{\,\cB}}.
$$
Then, the total Legendre transform of $\Psi_{\cZ_{0}}$ is defined as 
$$
\wt{\psi}^{\,*}_{\,\cZ_{0}}
=\int_{\cZ_{\,0}}\Leg[\psi]\,\rho^{\,d_{\cB}},
$$
where $\Leg[\psi]$ is the total Legendre transform of $\psi$. 
\end{Def}

As shown in 
Propositions\,\ref{Mrugala-variant-psi} and \ref{Mrugala-variant-varphi},
vector fields on Legendre submanifolds of contact manifolds are concisely 
written as contact Hamiltonian vector fields with adapted functions introduced 
in Definition\,\ref{definition-adapted-function} 
for the standard contact geometry. 
Also, for contact geometry on fiber spaces,  
similar functions can be defined as follows. 

\begin{Def}
\label{definition-adapted-function-on-bundle}
(Adapted functions on fiber space):  
Let $(\cK,\lambda_{\,\mbbV},\pi,\cB)$ be a $(2n+1)$-dimensional contact manifold
over a base space $\cB$, $(x,p,z)$ 
canonical coordinates for $\pi^{-1}(\zeta),(\zeta\in\cB)$ such that 
$\lambda_{\,\mbbV}=\dr_{\,\mbbV} z-p_{\,a}\,\dr_{\mbbV} x^{\,a}$ with 
$x=\{\,x^{\,1},\ldots,x^{\,n}\,\}$ and $p=\{\,p_{\,1},\ldots,p_{\,n}\,\}$, 
and $\cK=\bigcup_{\,\zeta\in\cB}\cC_{\,\zeta}$. 
In addition let 
$\psi$ be a vertical 
function on $\cC_{\,\zeta}$ depending on $x$,    
and $\varphi$ a vertical function on $\cC_{\,\zeta}$ depending on $p$. Then
the functions 
$\Delta_{\,0}^{\,\zeta\psi},\{\Delta_{\,1}^{\,\zeta\psi},\ldots,\Delta_{\,n}^{\,\zeta\psi}\}:\cC_{\,\zeta}\to\mbbR,$ 
and 
$\Delta_{\,\zeta\varphi}^{\,0},\{\Delta_{\,\zeta\varphi}^{\,1},\ldots,\Delta_{\,\zeta\varphi}^{\,n}\}:\cC_{\,\zeta}\to\mbbR$  
such that  
$$
\Delta_{\,0}^{\,\zeta\psi}(x,z)
:=\psi(x)-z,\quad 
\Delta_{\,a}^{\,\zeta\psi}(x,p)
:=\frac{\partial\psi}{\partial x^{\,a}}-p_{\,a},\qquad 
a\in\{\,1,\ldots,n\,\}.
$$
$$
\Delta_{\,\zeta\varphi}^{\,0}(x,p,z)
:=x^{\,j}p_{\,j}-\varphi(p)-z,\quad 
\Delta_{\,\zeta\varphi}^{\,a}(x,p)
:=x^{\,a}-\frac{\partial\varphi}{\partial p_{\,a}},\qquad 
a\in\{\,1,\ldots,n\,\}.
$$  
are referred to as adapted functions on the fiber space. 
\end{Def}

Similar to 
Proposition\,\ref{proposition-Legendre-submanifold-adapted-functions},
one has the following.
\begin{Proposition}
\label{proposition-Legendre-submanifold-adapted-functions-on-fiber}
(Local expressions of Legendre submanifold with adapted functions):  
The Legendre submanifold $\cA_{\,\psi}$ generated by $\psi$ in $\pi^{-1}(\zeta)$ 
as 
\fr{example-psi-Legendre-submanifold-bundle}
is expressed as 
\beq
\cA_{\,\zeta\psi}^{\,\cC}
:=\Phi_{\,\cC_{\,\zeta}\cA_{\,\zeta\psi}}\cA_{\,\zeta\psi}
=\left\{\ (x,p,z)\in\cC_{\zeta} \ |\ 
\Delta_{\,0}^{\,\zeta\psi}
=0\ \mbox{and}\ 
\Delta_{\,1}^{\,\zeta\psi}
=\cdots
=\Delta_{\,n}^{\,\zeta\psi}
=0
\ \right\}, 
\label{Legendre-submanifold-fiber-psi}
\eeq
where 
$\Phi_{\,\cC_{\,\zeta}\cA_{\,\zeta\psi}}\cA_{\,\zeta\psi}:\cA_{\,\zeta\psi}\to\cC_{\,\zeta}$ 
is the embedding. Similarly, 
the Legendre submanifold $\cA_{\,\varphi}$ generated by $-\varphi$ 
in $\pi^{\,-1}(\zeta)$ as 
\fr{example-varphi-Legendre-submanifold-bundle}
is expressed as 
\beq
\cA_{\,\zeta\varphi}^{\,\cC}
:=\Phi_{\,\cC_{\,\zeta}\cA_{\,\zeta\varphi}}\cA_{\,\zeta\varphi}
=\left\{\ (x,p,z)\in\cC_{\,\zeta} \ |\ 
\Delta_{\,\zeta\varphi}^{\,0}
=0\ \mbox{and}\ 
\Delta_{\,\zeta\varphi}^{\,1}
=\cdots
=\Delta_{\,\zeta\varphi}^{\,n}
=0
\ \right\}, 
\label{Legendre-submanifold-fiber-varphi}
\eeq
where 
$\Phi_{\,\cC_{\zeta}\cA_{\,\zeta\varphi}}\cA_{\,\zeta\varphi}:\cA_{\,\zeta\varphi}\to\cC_{\,\zeta}$ 
is the embedding. 
\end{Proposition} 

Contact Hamiltonian vertical vector fields are 
also written in terms of adapted functions on fiber spaces.

\begin{Proposition} 
\label{Mrugala-variant-psi-bundle}
(Restricted contact Hamiltonian vertical vector field  
as the push-forward of a vector field on 
the Legendre submanifold generated by $\psi$):  
Let $\{\,F_{\,\psi}^{\,\zeta,1},\ldots,F_{\,\psi}^{\,\zeta,n}\,\}$ 
be a set of functions of $x$ on $\cA_{\,\zeta\psi}$ 
such that they do not identically vanish,   
and 
$\chX_{\,\zeta\psi}^{\,0}\in T_{\,x}\,\cA_{\,\zeta\psi}, ( x\in \cA_{\,\zeta\psi})$ 
the vector field given as   
$$
\chX_{\,\zeta\psi}^{\,0}
=\dot{x}^{\,a}\,\frac{\partial}{\partial x^{\,a}},
\quad\mbox{where}\quad 
\dot{x}^{\,a}
=F_{\,\zeta\psi}^{\,a}(x),\qquad 
(a\in\{\,1,\ldots,n\,\}).
$$
In addition, let 
$X_{\,\zeta\psi}^{\,0}
:=(\,\Phi_{\,\cC_{\zeta}\cA_{\,\zeta\psi}}\,)_{*}\chX_{\,\zeta\psi}^{\,0} 
\in T_{\,\xi}\cA_{\,\zeta\psi}^{\,\cC}, (\,\xi\in\cA_{\,\zeta\psi}^{\,\cC}\,)$ 
be the push-forward of
$\chX_{\,\zeta\psi}^{\,0}$,   
where $\cA_{\,\zeta\psi}^{\,\cC}:=\Phi_{\,\cC_{\,\zeta}\cA_{\,\zeta\psi}}\cA_{\,\zeta\psi}$ with   
$\Phi_{\,\cC_{\,\zeta}\cA_{\,\zeta\psi}}:\cA_{\,\zeta\psi}\to \cC_{\,\zeta}$ 
being the embedding :  
\beqa
\Phi_{\,\cC_{\,\zeta}\cA_{\,\zeta\psi}} &:& \cA_{\,\zeta\psi}\to 
\cA_{\,\zeta\psi}^{\,\cC},\qquad\qquad x\mapsto (\,x,p(x),z(x)\,),\quad
\mbox{on}\quad \pi^{-1}(\zeta),
\non\\
(\,\Phi_{\,\cC_{\,\zeta}\cA_{\,\zeta\psi}}\,)_{\,*} 
&:& T_{\,x}\,\cA_{\,\zeta\psi}\to  
T_{\,\xi}\,\cA_{\,\zeta\psi}^{\,\cC},\quad \chX_{\,\zeta\psi}^{\,0}\mapsto
X_{\,\zeta\psi}^{\,0}\,.
\non
\eeqa
Then it follows that   
\beq
X_{\,\zeta\psi}^{\,0}
=\dot{x}^{\,a}\frac{\partial}{\partial x^{\,a}}
+\dot{p}_{\, a}\frac{\partial}{\partial p_{\, a}}
+\dot{z}\frac{\partial}{\partial z},\quad\mbox{where}\quad 
\dot{x}^{\,a}
=F_{\,\zeta\psi}^{\,a}(x),\quad
\dot{p}_{\,a}
=\frac{\dr}{\dr t}\left(\,\frac{\partial\psi}{\partial x^{\,a}}\,
\right),
\quad 
\dot{z}
=\frac{\dr \psi}{\dr t}.
\label{tangent-vector-Legendre-submanifold-psi-component-bundle}
\eeq
In addition, 
one has that $X_{\,\zeta\psi}^{\,0}=X_{\,\wt{h}_{\,\psi}}|_{\,\wt{h}_{\,\psi}=0}$. Here 
$X_{\,\wt{h}_{\,\psi}}$ is the contact Hamiltonian vertical vector field associated 
with  
\beq
h_{\,\psi}(x,p,z)
=\Delta_{\,a}^{\,\zeta\psi}(x,p) F_{\,\zeta\psi}^{\,a}(x) 
+\Gamma_{\,\zeta\psi}(\,\Delta_{\,0}^{\,\zeta\psi}(x,z)\,),
\label{tangent-vector-Legendre-submanifold-psi-Hamiltonian-bundle}
\eeq
where $\Gamma_{\,\zeta\psi}$ is a function of $\Delta_{\,0}^{\,\zeta\psi}$ such that 
$$
\Gamma_{\,\zeta\psi}\left(\,\Delta_{\,0}^{\,\zeta\psi}\,\right)
=\left\{
\begin{array}{cl}
0&\mbox{for}\quad \Delta_{\,0}^{\,\zeta\psi}=0\\
\mbox{non-zero}&\mbox{for}\quad\Delta_{\,0}^{\,\zeta\psi}\neq 0
\end{array}
\right..
$$  
\end{Proposition}
There exists 
a counterpart of Proposition\,\ref{Mrugala-variant-psi-bundle}, that 
is given as follows.
 
\begin{Proposition} 
\label{Mrugala-variant-varphi-bundle}
(Restricted contact Hamiltonian vertical vector field  
as the push-forward of a vector field on 
the Legendre submanifold generated by $-\varphi$):  
Let $\{\,F_{\,\zeta,1}^{\,\varphi},\ldots,F_{\,\zeta,n}^{\,\varphi}\,\}$ 
be a set of functions of $p$ on $\cA_{\,\zeta\varphi}$ 
such that they do not identically vanish,   
and 
$\chX_{\,0}^{\,\zeta\varphi}\in T_{\,p}\,\cA_{\,\zeta\varphi}, ( p\in \cA_{\,\zeta\varphi})$ 
the vector field given as   
$$
\chX_{\,0}^{\,\zeta\varphi}
=\dot{p}_{\,a}\,\frac{\partial}{\partial p_{\,a}},
\quad\mbox{where}\quad 
\dot{p}_{\,a}
=F_{\,a}^{\,\zeta\varphi}(p),\qquad 
(a\in\{\,1,\ldots,n\,\}).
$$
In addition, let 
$X_{\,0}^{\,\zeta\varphi}
:=(\,\Phi_{\,\cC_{\zeta}\cA_{\,\zeta\varphi}}\,)_{*}\chX_{\,0}^{\,\zeta\varphi} 
\in T_{\,\xi}\cA_{\,\zeta\varphi}^{\,\cC}, (\,\xi\in\cA_{\,\zeta\varphi}^{\,\cC}\,)$ 
be the push-forward of
$\chX_{\,0}^{\,\zeta\varphi}$,   
where $\cA_{\,\zeta\varphi}^{\,\cC}:=\Phi_{\,\cC_{\,\zeta}\cA_{\,\zeta\varphi}}\cA_{\,\zeta\varphi}$
 with $\Phi_{\,\cC_{\,\zeta}\cA_{\,\zeta\varphi}}:\cA_{\,\zeta\varphi}\to \cC_{\,\zeta}$ 
being the embedding :  
\beqa
\Phi_{\,\cC_{\,\zeta}\cA_{\,\zeta\varphi}} &:& \cA_{\,\zeta\varphi}\to 
\cA_{\,\zeta\varphi}^{\,\cC},\qquad\qquad p\mapsto (\,x(p),p,z(p)\,),\quad
\mbox{on}\quad \pi^{-1}(\zeta),
\non\\
(\,\Phi_{\,\cC_{\,\zeta}\cA_{\,\zeta\psi}}\,)_{\,*} 
&:& T_{\,p}\,\cA_{\,\zeta\varphi}\to  
T_{\,\xi}\,\cA_{\,\zeta\varphi}^{\,\cC},\quad \chX_{\,0}^{\,\zeta\varphi}
\mapsto
X_{\,0}^{\,\zeta\varphi}\,.
\non
\eeqa
Then, it follows that   
\beq
X_{\,0}^{\,\zeta\varphi}
=\dot{x}^{\,a}\frac{\partial}{\partial x^{\,a}}
+\dot{p}_{\, a}\frac{\partial}{\partial p_{\, a}}
+\dot{z}\frac{\partial}{\partial z},\ \mbox{where}\quad  
\dot{x}^{\,a}
=\frac{\dr}{\dr t}\left(\,\frac{\partial\varphi}{\partial p_{\,a}}\,
\right),\quad
\dot{p}_{\,a}
=F_{\,a}^{\,\zeta\varphi}(p)
,\quad 
\dot{z}
=p_{\,j}F_{\,k}^{\,\zeta\varphi}
\frac{\partial^2 \varphi}{\partial p_{\,k}\partial p_{\,j} }.
\label{tangent-vector-Legendre-submanifold-varphi-component-bundle}
\eeq
In addition, 
one has that $X_{\,0}^{\,\zeta\varphi}=X_{\,\wt{h}_{\,\varphi}}|_{\,\wt{h}_{\,\varphi}=0}$. Here 
$X_{\,\wt{h}_{\,\varphi}}$ is the contact Hamiltonian vertical vector field 
associated with  
\beq
h_{\,\varphi}(x,p,z)
=\Delta_{\,\zeta\varphi}^{\,a}(x,p) F_{\,a}^{\,\zeta\varphi}(p) 
+\Gamma^{\,\zeta\varphi}(\,\Delta_{\,\zeta\varphi}^{\,0}(x,p,z)\,),
\label{tangent-vector-Legendre-submanifold-varphi-Hamiltonian-bundle}
\eeq
where $\Gamma^{\,\zeta\varphi}$ is a function of $\Delta_{\,\zeta\varphi}^{\,0}$ 
such that 
$$
\Gamma^{\,\zeta\varphi}\left(\,\Delta_{\,\zeta\varphi}^{\,0}\,\right)
=\left\{
\begin{array}{cl}
0&\mbox{for}\quad \Delta_{\,\zeta\varphi}^{\,0}=0\\
\mbox{non-zero}&\mbox{for}\quad\Delta_{\,\zeta\varphi}^{\,0}\neq 0
\end{array}
\right..
$$  
\end{Proposition}

\section{Maxwell's equations without source }
In this section the $(3+1)$-decomposed  
Maxwell's equations without source in terms of a form language is summarized.   
This formulation is standard and can be found in the literature\cite{Frankel}. 

\subsection{Three-dimensional Riemannian manifold}
To discuss $(3+1)$-decomposed Maxwell's equations in our extended framework of 
contact geometry,  
a bundle will be used. 
In this extended framework, the base space is 
a $3$-dimensional Riemannian manifold.  

Let $(\cZ,g)$ be a $3$-dimensional Riemannian manifold, and $\star 1$ 
the canonical volume form, 
$\star:\Gamma\Lambda^{\,q}\cZ\to\Gamma\Lambda^{\,3-q}\cZ,(q\in\{0,\ldots,3\})$ 
the Hodge dual map : 
$$
\star(\,\balpha\wedge \bgamma\,)
=\ii_{\,Y}\star\,\balpha,\quad
\star(\,f\,\balpha\,)
=f\,\star\,\balpha,\quad
\star(\,\balpha+\bbeta\,)
=\star\balpha+\star\bbeta,\qquad 
$$  
where $Y\in\Gamma T\cZ$ is such that $\bgamma=g(\,Y,-)$,
for all $\balpha,\bbeta\in\Gamma\Lambda^{\,q}\cZ,\bgamma\in\Gamma\Lambda^{\,1}\cZ,f\in\Gamma\Lambda^{\,0}\cZ, (q\in\{0,\ldots,3\})$. In addition,  
let $\{\bsigma^{\,a}\}$ be the set of orthogonal co-frames being dual to 
$\{\,X_{\,a}\,\}$ 
so that $g=\delta_{\,ab}\,\bsigma^{\,a}\otimes\bsigma^{\,b}$, 
$\bsigma^{\,a}(X_{\,b})=\delta_{\,ab}$ and 
$\star1=\bsigma^{\,1}\wedge \bsigma^{\,2}\wedge \bsigma^{\,3}$. 
Then, the contravariant metric tensor field is 
$g^{-1}=\delta_{\,ab}X^{\,a}\otimes X^{\,b}$, and 
$\star(\bsigma^{\,b}\wedge\bsigma^{\,c})=\epsilon_{\,a}^{\ \ bc}\bsigma^{\,a}$,
where
$$
\epsilon^{\, abc}
=\epsilon_{\,a}^{\ \ bc}
=\left\{
\begin{array}{cl}
+1&\mbox{even permutation of $a=1,b=2,c=3$}\\
-1&\mbox{odd permutation of $a=1,b=2,c=3$}\\
0&\mbox{other}.
\end{array}
\right. 
$$

The following will be used.
\begin{Lemma}
For any $\balpha,\bbeta\in\Gamma\Lambda^{\,q}{\cZ}$ with $q\in\{0,\ldots,3\}$, 
it follows that 
$$
\star\star\,\balpha
=\balpha,\quad\mbox{and}\quad 
\balpha\wedge\star\bbeta
=\bbeta\wedge\star\balpha.
$$
\end{Lemma}
\begin{Lemma}
For any $1$-form $\bDelta=\Delta_{\,a}\bsigma^{\,a}$ and $2$-form 
$\bF=(1/2)\,F_{\,ab}\,\bsigma^{\,a}\wedge\bsigma^{\,b}$ with $F_{\,ab}=-\,F_{ba}$, 
it follows that 
\beq  
\star\left(\,\bDelta\wedge\bF\,\right)
=g^{\,-1}(\bDelta,\star\bF)
=\delta^{\,ab}\Delta_{\,a}(\star\bF)_{\,b}
=\Delta_{\,1} F_{23}+\Delta_{\,2} F_{31}+\Delta_{\,3} F_{12}, 
\label{Hodge-2-1}
\eeq
where 
$(\star\bF)_{\,a}$ is such that $\star\bF=(\star\bF)_{\,a}\,\bsigma^{\,a}$.
\end{Lemma} 

Some examples of the functional derivative 
introduced in Definition\,\ref{definition-functional-derivative}
are shown below.  
\begin{Example} 
Let $\alpha$ be a $0$-form, and $\psi$ a function of $\alpha$. 
Consider the functional 
$$
\wt{\psi}_{\cZ_{\,0}}\left[\,\alpha\,\right]
=\int_{\cZ_{\,0}} \psi(\alpha)\,\star 1.
$$
Then, 
$$
\frac{\delta\,\wt{\psi}_{\cZ_{\,0}}}{\delta\alpha}
=\frac{\partial \psi}{\partial \alpha}.
$$
\end{Example}
\begin{Example} 
Let $\balpha$ be a $1$-form written as $\balpha=\alpha_{\,a}\,\bsigma^{\,a}$.  
Consider the functional
$$
\wt{\psi}_{\cZ_{\,0}}\left[\,\balpha\,\right]
=\frac{1}{2}\int_{\cZ_{\,0}}\balpha\wedge\star\balpha
=\int_{\cZ_{\,0}}\psi\,\star 1,\qquad\mbox{where}\qquad 
\psi=\frac{1}{2}\,g^{-1}(\balpha,\balpha)
=\frac{1}{2}\,\delta^{\,ab}\alpha_{\,a}\alpha_{\,b}.
$$
Then, 
$$
\frac{\delta\,\wt{\psi}_{\cZ_{\,0}}}{\delta\balpha}
=\star\balpha,\qquad\mbox{and}\qquad
\frac{\partial \psi}{\partial\alpha_{\,a}}
=\delta^{\,ab}\alpha_{\,b}. 
$$
\end{Example}
\begin{Example} 
\label{example-functional-derivative-2}
Let $\bbeta$ be a $2$-form written as 
$\bbeta=(1/2)\beta_{\,ab}\,\bsigma^{\,a}\wedge\bsigma^{\,b}$, and 
$\star\bbeta=(\star\bbeta)_{\,a}\bsigma^{\,a}$.  
Consider 
$$
\wt{\psi}_{\cZ_{\,0}}\left[\,\bbeta\,\right]
=\frac{1}{2}\int_{\cZ_{\,0}}\bbeta\wedge\star\bbeta
=\int_{\cZ_{\,0}}\psi\,\star 1,\qquad\mbox{where}\qquad
\psi=\frac{1}{2}\,g^{-1}\left(\star\bbeta,\star\bbeta\right).
$$
Then, 
$$
\frac{\delta\,\wt{\psi}_{\cZ_{\,0}}}{\delta\bbeta}
=\star\bbeta,\qquad\mbox{and}\qquad
\frac{\partial \psi}{\partial\beta^{\,a}}
=\delta_{\,ab}\beta^{\,b}, 
$$
where 
$$
\beta^{\,a}
:=\delta^{\,ab}(\star\bbeta)_{\,b},\quad\mbox{so that}\quad
\psi=\frac{1}{2}\,\delta_{\,ab}\,\beta^{\,a}\,\beta^{\,b}.
$$
\end{Example}

\subsection{Maxwell fields}
The forms  
$\bfe,\bfh\in\GamLam{1}{\cZ}$ and $\bfD,\bfB\in\GamLam{2}{\cZ}$ are used 
for describing Maxwell's equations. 
Their physical meanings are given as below :  
$$
\begin{array}{ll}
\bfe&:\ \mbox{$1$-form electric field}\\
\bfB&:\ \mbox{$2$-form magnetic induction field}\\
\bfD&:\ \mbox{$2$-form displacement field}\\
\bfh&:\ \mbox{$1$-form magnetic field}
\end{array}
$$


With 
$\bfe,\bfh\in\GamLam{1}{\cZ}$ and $\bfD,\bfB\in\GamLam{2}{\cZ}$, 
Maxwell's equations without source are written as follows. 
\begin{Def}
(Maxwell's equations without source): 
The  $(3+1)$-decomposed Maxwell's equations without external source are 
\beq
\dot{\bfD}
=\dr\bfh,\qquad 
\dot{\bfB}
=-\,\dr\bfe,\qquad 
\dr \bfD
=0,\qquad 
\dr \bfB
=0.
\label{decomposed-Maxwell-without-source-media}
\eeq
Here $\dot{\,}$ is derivative with respect to time. 
\end{Def}
\begin{Remark}
The two equations, $\dr\bfD=0$ and $\dr\bfB=0$, 
can be derived from the other two equations, $\dot{\bfD}=\dr \bfh$ and $\dot{\bfB}=-\,\dr\bfe$, by 
applying $\dr$ with $\dr^2=0$.
\end{Remark}

To obtain closed equations from Maxwell's equations without source, 
one needs some relations. 
\begin{Def}
(Constitutive relation): The following relations 
\begin{enumerate}
\item  
a relation connecting $\bfe$ with  $\bfD,\bfB$ 
\item
a relation connecting $\bfh$ with $\bfB,\bfD$ 
\end{enumerate}
are referred to as constitutive relations.
\end{Def} 
There are special constitutive relations with some functions on $\cZ$ 
and on possibly time $t$.  
The following are the typical ones.
\begin{Def}
\label{definition-permittivity-permeability}
(Permittivity and permeability):  
For Maxwell's equations, if $\bfD=\ve\,\star\bfe$ and $\bfB=\mu\star\bfh$ 
with some positive $\ve$ and $\mu$ that do not depend on $\bfe,\bfB,\bfD,\bfh$, 
then $\ve$ is referred to as permittivity,
and $\mu$ as permeability, respectively. 
\end{Def} 
\begin{Remark}
If $\ve$ or $\mu$ depends on $\bfe,\bfB,\bfD,\bfh$, then  
Maxwell's equations are nonlinear. 
In this paper linear case is only considered.
\end{Remark}
\begin{Remark}
Consider the case where $\dot{\ve}=\dot{\mu}=0$ on some space on $\cZ$.  
In this case, one can introduce a potential $1$-form $\bfA$ such that 
$\bfB=\dr \bfA$. It then follows that $\bfe=-\dot{\bfA}$, and one has 
the equation  of motion for $\bfA$ 
$$
\ddot{\bfA}+\frac{1}{\ve}\,\star\dr\left(\,
\frac{1}{\mu}\star \dr \bfA
\,\right)
=0.  
$$
From this, one can consider various electromagnetic systems. 
For example, 
choose  
$\ve(\zeta_3)=\ve_{0}\,\sech^{\,2}(\zeta_{\,3}/\zeta_{\,30})$ and $\mu=\mu_{\,0}$ 
where $\zeta_{\,3}$ is a coordinate for $\cZ$ and 
$\ve_{\,0},\zeta_{\,30},\mu_{\,0}$ constants. Then one has 
analytical expressions of electromagnetic waves for this model\cite{GTW2014}.
In addition choose $\ve=\ve_{\,0}$ and $\mu=\mu_{\,0}$ with $\ve_{\,0}$ 
and $\mu_{\,0}$ being constants. Then one has the vacuum system. 
In this case there exist various solutions whose field lines 
form knots\cite{Arrayas}.
\end{Remark}
Although there are a variety of constitutive relations,   
this special class of constitutive relations, involving 
$\ve$ and $\mu$ as in Definition\,\ref{definition-permittivity-permeability}, 
are only considered in this paper since they are typical and 
mathematically simple.

The forms  $\bfe,\bfB,\bfD,\bfh$ are classified as follows.
\begin{Def}
(Maxwell fields, induction field and field intensity): 
The forms $\bfe,\bfB,\bfD,\bfh$ that are used to describe Maxwell's equations 
are referred to as the Maxwell fields. In addition,  
the $2$-forms $\bfD$ and $\bfB$ are referred to as induction fields, 
and the $1$-forms $\bfe$ and $\bfh$ are referred to as field intensities.  
\end{Def}
\begin{Remark}
There exists another classification for these forms\cite{Frankel}. The forms 
$\bfe$ and $\bfB$ are untwisted forms, and $\bfD$ and $\bfh$ 
twisted forms. 
\end{Remark}

Given a medium with given boundary, the forms should satisfy 
appropriate boundary conditions. 
Taking into account this, one assumes that the solutions to 
Maxwell's equations always satisfy such boundary conditions in discussions 
below.

In physics energy plays a role, and energy functionals are 
used for continuous mechanics.  In electromagnetism, 
the following functionals can be chosen and 
used in this paper. 
\begin{Def}
(Energy and co-energy functionals): 
Let $\bfe,\bfB,\bfD,\bfh$ be solutions to Maxwell's equations without source, 
$\ve:\cZ\to\mbbR$ and $\mu:\cZ\to\mbbR$ permittivity and 
permeability depending on at a point $\zeta$ of $\cZ$ and time $t\subset\mbbT$, 
respectively, and 
$\cZ_{\,0}\subseteq\cZ$ a subspace of $\cZ$.  
The functional 
\beq
\wt{\psi}_{\cZ_{\,0}}^{\,\EM}\left[\bfD,\bfB\right]
:=\frac{1}{2}\int_{\cZ_{\,0}}\left(
\frac{1}{\ve}\bfD\wedge\star\bfD
+\frac{1}{\mu}\bfB\wedge\star\bfB\right),
\label{Hamiltonian-D-B}
\eeq
is referred to as the energy functional. 
In addition, 
\beq
\wt{\varphi}_{\cZ_{\,0}}^{\,\EM}\left[\bfe,\bfh\right]
:=\frac{1}{2}\int_{\cZ_{\,0}}\left(\,\ve\,\bfe\wedge\star\bfe
+\mu\,\bfh\wedge\star\bfh\right),
\label{Hamiltonian-e-h}
\eeq
is referred to as the co-energy functional.
\end{Def}
The functional  $\wt{\psi}_{\cZ_{\,0}}^{\,\EM}$ in \fr{Hamiltonian-D-B}
depends on induction fields, and this 
will lead to 
the induction oriented formulation of Maxwell's equations.
On the other hand, 
$\wt{\varphi}_{\cZ_{\,0}}^{\,\EM}$ in \fr{Hamiltonian-e-h} 
depends on field intensities, and this 
will lead to 
the field intensity oriented formulation of Maxwell's equations.

The decomposed Maxwell's equations are written on 
a $3$-dimensional Riemannian manifold $(\cZ,g)$. 
With the Riemannian metric tensor field $g$ one can write  
\fr{Hamiltonian-D-B} and \fr{Hamiltonian-e-h} as follows. 
\begin{Lemma}
The functionals \fr{Hamiltonian-D-B} and \fr{Hamiltonian-e-h} can be 
written as 
$$
\wt{\psi}_{\,\cZ_{\,0}}^{\,\EM}\left[\bfD,\bfB\right]
=\int_{\cZ_{0}}\psi^{\,\EM}(\bfD,\bfB)\star 1,\qquad\mbox{and}\qquad
\wt{\varphi}_{\,\cZ_{\,0}}^{\,\EM}\left[\bfe,\bfh\right]
=\int_{\cZ_{0}}\varphi^{\,\EM}(\bfe,\bfh)\star 1,
$$
where 
\beq
\psi^{\,\EM}(\bfD,\bfB)
=\frac{1}{2}\left[\,\frac{1}{\ve}\,g^{-1}(\star\bfD,\star\bfD)
+\frac{1}{\mu}\,g^{-1}(\star\bfB,\star\bfB)\,\right],\quad 
\varphi^{\,\EM}(\bfe,\bfh)
=\frac{1}{2}\left[\,\ve\, g^{-1}(\bfe,\bfe)+\mu\, g^{-1}(\bfh,\bfh)\,
\right].
\label{definition-energy-density-function}
\eeq
\end{Lemma}
\begin{Proof}
With the identities
$$
\frac{1}{\ve}\,\bfD\wedge\star\bfD
+\frac{1}{\mu}\,\bfB\wedge\star\bfB
=\left[\,\frac{1}{\ve}\,g^{-1}(\star\bfD,\star\bfD)
+\frac{1}{\mu}\,g^{-1}(\star\bfB,\star\bfB)
\,\right]\,\star 1,
$$
and 
$$
\ve\,\bfe\wedge\star\bfe
+\mu\,\bfh\wedge\star\bfh
=\left[\,\ve\, g^{-1}(\bfe,\bfe)
+\mu\, g^{-1}(\bfh,\bfh)
\,\right]\,\star 1,
$$
one can complete the proof.
\qed
\end{Proof}
\begin{Def}
(Energy density function and co-energy density function): 
The functions $\psi^{\,\EM}$ and 
$\varphi^{\,\EM}$ 
in \fr{definition-energy-density-function}
are referred to as the energy density function and 
the co-energy density function, respectively. 
\end{Def}

In the following, various formulae are shown. 
They will be used in the rest of this section 
and 
Section\,\ref{section-Maxwell-equation-contact-formulation}.  
 
 
\begin{Lemma}
\label{derivative-Psi-psi-D}
\beq
\frac{\delta\wt{\psi}_{\,\cZ_{\,0}}^{\,\EM}}{\delta\bfD}
=\frac{1}{\ve}\star\bfD,\qquad 
\frac{\partial\psi^{\,\EM}}{\partial D^{\,a}}
=\frac{1}{\ve}(\star\bfD)_{a}
=\frac{1}{\ve}\delta_{\,ab}D^{\,b}
=\left(\frac{\delta\wt{\psi}_{\cZ_{0}}^{\,\EM}}{\delta\bfD}\right)_{\,a},
\label{derivative-Psi-psi-D-equations}
\eeq
where $D^{\,a}:=\delta^{\,ab}(\star\bfD)_{\,b}$, 
$$
\star\bfD
=(\star\bfD)_{\,a}\,\bsigma^{\,a},\qquad\mbox{and}\qquad
\frac{\delta\wt{\psi}_{\,\cZ_{\,0}}^{\,\EM}}{\delta\bfD}
=\left(\frac{\delta\wt{\psi}_{\cZ_{\,0}}^{\,\EM}}{\delta\bfD}\right)_{\,a}
\,\bsigma^{\,a}.
$$
\end{Lemma}
\begin{Proof}
With Example\,\ref{example-functional-derivative-2}, 
one can easily prove the first equation.  
In the following, the second equation is proven.  
One can write $\bfD$ with $D_{\,ab}=-D_{\,ba}$ as 
$$
\bfD=\frac{1}{2}D_{\,bc}\,\bsigma^{\,b}\wedge \bsigma^{\,c}.
$$
It is straightforward to show that 
$$
\star\bfD
=\frac{1}{2}D_{\,bc}\,\star\left(\,\bsigma^{\,b}\wedge \bsigma^{\,c}\,\right)
=\frac{1}{2}\epsilon_{\,a}^{\ \ bc}\,D_{\,bc}\,\bsigma^{\,a},
$$
and 
$$
(\star\bfD)_{\,a}
=(\star\bfD)(X_{\,a})
=\frac{1}{2}\epsilon_{\,a}^{\ \ bc}\,D_{\,bc},\qquad\mbox{and}\qquad
D^{\,a}
=\delta^{\,ab}(\star\bfD)_{\,b}
=\frac{\epsilon^{\,abc}}{2}D_{\,bc}.
$$
Then, it follows from 
$$
g^{-1}(\star\bfD,\star\bfD)
=\delta^{aa'}\left(\,\frac{1}{2}\epsilon_{\,a}^{\ \ bc}D_{\,bc}\,\right)\,
\left(\frac{1}{2}\epsilon_{\,a'}^{\ \ b'c'}D_{\,b'c'}\right)
=\delta_{\,aa'}\left(\,\frac{1}{2}\epsilon^{\ abc}D_{\,bc}\,\right)\,
\left(\frac{1}{2}\epsilon^{\ a'b'c'}D_{\,b'c'}\right)
=\delta_{\,aa'}\,D^{\,a}D^{\,a'},
$$ 
that 
$$
\frac{\partial\psi^{\,\EM}}{\partial D^{\,a}}
=\frac{1}{2\,\ve}\frac{\partial}{\partial D^{\,a}}
\left[\,
g^{-1}(\star\bfD,\star\bfD)
\,\right]
=\frac{1}{2\,\ve}\frac{\partial}{\partial D^{\,a}}
\left[\,
\delta_{\,bc}\,D^{\,b}D^{\,c}
\,\right]
=\frac{1}{\ve}\delta_{\,ab}D^{\,b}
=\frac{1}{\ve}(\star\bfD)_{\,a}.
$$
\qed
\end{Proof}

Similar to Lemma\,\ref{derivative-Psi-psi-D},  one has the following.
\begin{Lemma}
\label{derivative-Psi-psi-B}
\beq
\frac{\delta\wt{\psi}_{\,\cZ_{\,0}}^{\,\EM}}{\delta\bfB}
=\frac{1}{\mu}\star\bfB,\qquad 
\frac{\partial\psi^{\,\EM}}{\partial B^{\,a}}
=\frac{1}{\mu}(\star\bfB)_{a}
=\frac{1}{\mu}\delta_{\,ab}B^{\,b}
=\left(\frac{\delta\wt{\psi}_{\,\cZ_{\,0}}^{\,\EM}}{\delta\bfB}\right)_{\,a},
\label{derivative-Psi-psi-B-equations}
\eeq
where $B^{\,a}:=\delta^{\,ab}(\star\bfB)_{\,b}$, 
$$
\star\bfB
=(\star\bfB)_{\,a}\,\bsigma^{\,a},\qquad\mbox{and }\qquad
\frac{\delta\wt{\psi}_{\,\cZ_{\,0}}^{\,\EM}}{\delta\bfD}
=\left(\frac{\delta\wt{\psi}_{\cZ_{\,0}}^{\,\EM}}{\delta\bfD}\right)_{\,a}
\,\bsigma^{\,a}.
$$
\end{Lemma}
\begin{Proof}
A way to prove this is analogous to the proof of  
Lemma\,\ref{derivative-Psi-psi-D}.
\qed
\end{Proof}

With the formulae derived above, one can characterize the energy density 
function and the co-energy density function. To this end, one defines 
convex function for smooth functions.
\begin{Def}
(Strictly convex function): 
Let $\cA_{\,0}\subseteq \cA$ be a convex domain, and 
$f$ a function of $\{x^{\,a}\}$ on $\cA_{\,0}$. If the matrix  
$$
\frac{\partial^2\, f}{\partial x^{\,a}\partial x^{\,b}}
$$
is strictly positive definite, then the function is referred to as  
a strictly convex function. 
In addition, the property that $f$ is strictly convex is referred to 
as convexity. 
\end{Def}
From this definition, 
Lemma\,\ref{derivative-Psi-psi-D}, and  
Lemma\,\ref{derivative-Psi-psi-B}, one has the following.  
\begin{Lemma}
\label{convex-psi-varphi}
(Convexity for $\psi^{\,\EM}$ and $\varphi^{\,\EM}$): The functions 
$\psi^{\,\EM}$ and $\varphi^{\,\EM}$ 
in \fr{definition-energy-density-function} 
are strictly convex functions.
\end{Lemma}
\begin{Proof}
Define 
$\{x^{\,a}\}=\{D^{\,1},D^{\,2},D^{\,3},B^{\,1},B^{\,2},B^{\,3}\}$ and 
$\{p_{\,a}\}=\{e_{\,1},e_{\,2},e_{\,3},h_{\,1},h_{\,2},h_{\,3}\}$.
It then follows from 
$\ve^{\,-1}>0,\mu^{\,-1}>0$ due to 
Definition\,\ref{definition-permittivity-permeability} 
that 
$$
\frac{\partial^{\,2}\psi^{\,\EM}}{\partial x^{\,a}\partial x^{\,b}}
=\diag{\ve^{\,-1},\ve^{\,-1},\ve^{\,-1},\mu^{\,-1},\mu^{\,-1},\mu^{\,-1}},
$$
which is strictly positive definite. 
Similarly one can prove that for $\varphi^{\,\EM}$.
\qed
\end{Proof}

The following states a relation between the energy density function and 
the co-energy density function.
\begin{Lemma}
\label{Legendre-transform-energy-density-function}
(Total Legendre transform of $\psi^{\,\EM}$):  
The functions $\psi^{\,\EM}$ and $\varphi^{\,\EM}$ defined 
in \fr{definition-energy-density-function} are related with  
the total Legendre transform : 
$$
\Leg[\psi^{\,\EM}]\left(\,\bfe,\bfh\,\right)
=\varphi^{\,\EM}\left(\,\bfe,\bfh\,\right).
$$
\end{Lemma}
\begin{Proof}
The total Legendre transform of $\Leg[\,\psi^{\,\EM}\,]$ is calculated to be 
\beqa
\Leg[\,\psi^{\,\EM}\,]\left(\,\bfe,\bfh\,\right)
&=&\sup_{\{D^{\,a}\},\{B^{\,a}\}}
\left[\,D^{\,a}e_{\,a}+B^{\,a}h_{\,a}
-\psi^{\,\EM}\,\right]
=D_{\,0}^{\,a}e_{\,a}+B_{\,0}^{\,a}h_{\,a}
-\psi^{\,\EM}(D_{\,0},B_{\,0})
\non\\
&=&\frac{1}{2\ve}\delta^{\,ab}e_{\,a}e_{\,b}
+\frac{1}{2\mu}\delta^{\,ab}h_{\,a}h_{\,b}
=\varphi^{\,\EM}(\bfe,\bfh),
\non
\eeqa
where we have used 
$D_{\,0}=\{D_{\,0}^{\,a}\}$ and $B_{\,0}=\{B_{\,0}^{\,a}\}$ that are the unique 
solutions to 
$$
e_{\,a}
=\frac{\partial\,\psi^{\,\EM}}{\partial D^{\,a}}(D_{\,0},B_{\,0}),
\qquad
h_{\,a}
=\frac{\partial\,\psi^{\,\EM}}{\partial B^{\,a}}(D_{\,0},B_{\,0}).
$$
The uniqueness follows from Lemma\,\ref{convex-psi-varphi}. 
\qed
\end{Proof}
 
The following states a relation between the two functionals.  
\begin{Lemma}
(Total Legendre transform of $\wt{\psi}_{\,\cZ_{0}}^{\,\EM}$): 
The total Legendre transform of the functional $\wt{\psi}_{\,\cZ_{0}}^{\,\EM}$ in 
\fr{Hamiltonian-D-B}
is $\wt{\varphi}_{\cZ_{0}}^{\,\EM}$ in \fr{Hamiltonian-e-h} : 
$$
\wt{\psi}_{\cZ_{0}}^{\,\EM\,*}[\,\bfe,\bfh\,]
=\wt{\varphi}_{\cZ_{0}}^{\,\EM}[\,\bfe,\bfh\,].
$$
\end{Lemma}
\begin{Proof}
From Definition\,\ref{definition-Legendre-transform-functional}, one has
$$
\wt{\psi}_{\,\cZ_{\,0}}^{\,\EM\,*}
=\int_{\cZ_{\,0}}\Leg[\,\psi^{\,\EM}\,]\star 1.
$$
Applying Lemma\,\ref{Legendre-transform-energy-density-function}, one has  
$$
\wt{\psi}_{\,\cZ_{0}}^{\,\EM\,*}[\,\bfe,\bfh\,]
=\int_{\cZ_{0}}\Leg[\,\psi^{\,\EM}\,]\star 1
=\int_{\cZ_{0}}\varphi^{\,\EM}(\bfe,\bfh)\,\star 1
=\wt{\varphi}_{\,\cZ_{0}}^{\,\EM}[\,\bfe,\bfh\,].
$$
\qed
\end{Proof}

\section{Contact formulation of Maxwell's equations without source}
\label{section-Maxwell-equation-contact-formulation}
In this section, from the given energy functional \fr{Hamiltonian-D-B} and 
given co-energy functional \fr{Hamiltonian-e-h},  
Maxwell's equations are formulated.

To this end, the field components of $\bfD,\bfB,\bfe,\bfh$, 
and the energy density function, 
$(\,D^{\,a},B^{\,a},e_{\,a},h_{\,a},\cE\,), (a\in\{1,2,3\})$,  
are identified with vertical $0$-forms that are defined as follows. 
\begin{Def}
(Maxwell fields and canonical coordinates of contact manifold):  
Let $\bfe,\bfD,\bfB,\bfh$ be the Maxwell fields on a Riemannian manifold 
$(\cZ,g)$. Then 
$D^{\,a},B^{\,a},e_{\,a},h_{\,a}$ are defined such that  
$$
D^{\,a}
:=\delta^{\,ab}(\star\bfD)_{\,b},\quad
B^{\,a}
:=\delta^{\,ab}(\star\bfB)_{\,b},\quad \mbox{and }\quad
\bfe=e_{\,a}\,\bsigma^{\,a},\quad
\bfh=h_{\,a}\,\bsigma^{\,a},\qquad a\in\{1,2,3\}.
$$
\end{Def}
\begin{Def}
(Contact manifold over a base space for Maxwell's equations):  
Let $(\cZ,g)$ be a $3$-dimensional Riemannian manifold, 
$(\cK,\pi,\cZ)$ a bundle over $\cZ$ and the fiber space $\pi^{-1}(\zeta)$ equals 
to a $13$-dimensional manifold $\cC_{\,\zeta}$ with 
$\cK=\bigcup_{\zeta\in\cZ}\cC_{\,\zeta}$, $(x^{\,\EM},p^{\,\EM},z^{\,\EM})$ 
canonical coordinates for the fiber space which are 
$$
x^{\,\EM}=\{D^1,D^2,D^3,B^1,B^2,B^3\},\qquad
p^{\,\EM}=\{e_1,e_2,e_3,h_1,h_2,h_3\},\qquad
z^{\,\EM}=\cE,
$$
with $\cE$ being either an energy density function or a co-energy density 
function, 
and $\lambda_{\,\mbbV}$ the following contact vertical form
$$
\lambda_{\mbbV}
=\dr_{\mbbV}z-p_a\dr_{\mbbV}x^{\,a}.
$$
The contact manifold over the base space $\cZ$ 
for the decomposed Maxwell's equations without source in 
media, 
\fr{decomposed-Maxwell-without-source-media}, 
is $(\cK,\lambda_{\mbbV},\pi,\cZ)$.  
\end{Def}

To close the Maxwell's equations, the constitutive relations 
\beq
e_{\,a}
=\frac{\partial\,\psi^{\,\EM}}{\partial D^{\,a}},
\qquad 
h_{\,a}
=\frac{\partial\,\psi^{\,\EM}}{\partial B^{\,a}},\quad a\in\{1,2,3\},
\label{constitutive-relation-Maxwell-e-h}
\eeq
or 
\beq
D^{\,a}
=\frac{\partial\,\varphi^{\,\EM}}{\partial e_{\,a}},
\qquad 
B^{\,a}
=\frac{\partial\,\varphi^{\,\EM}}{\partial h_{\,a}},\quad a\in\{1,2,3\},
\label{constitutive-relation-Maxwell-D-B}
\eeq
are imposed. 
In addition, one specifies the electromagnetic energy as either 
$$
\cE=\psi^{\,\EM}\qquad\mbox{or}\qquad 
\cE=D^{\,a}e_{\,a}+B^{\,a}h_{\,a}-\varphi^{\,\EM}.
$$
From a viewpoint of differential geometry, 
these relations are conditions that a solution space of Maxwell's equations 
is a Legendre submanifold generated by $\psi^{\,\EM}$ and that by 
$\varphi^{\,\EM}$.

The fiber space is a $13$-dimensional 
space that is for the expressing unrestricted fields and energy 
$(D^{\,a},B^{\,a},e_{\,a},h_{\,a},\cE)$,
and the Legendre submanifold is for expressing 
the restricted fields and energy.

\subsection{$\bfD$-$\bfB$ oriented formulation}
In this subsection from the given energy functional \fr{Hamiltonian-D-B}, 
Maxwell's equations will be formulated.

Impose $\cE=\psi^{\,\EM}$ and the constitutive relations 
\fr{constitutive-relation-Maxwell-e-h} 
$$
\bfe=\frac{\delta \wt{\psi}_{\,\cZ_{\,0}}^{\,\EM}}{\delta\bfD},
\qquad
\bfh=\frac{\delta \wt{\psi}_{\,\cZ_{\,0}}^{\,\EM}}{\delta\bfB},
$$
or equivalently 
$$
\bfe
=\frac{1}{\ve}\star\bfD,\qquad 
\bfh
=\frac{1}{\mu}\star\bfB,\qquad\mbox{or}\qquad
e_{\,a}
=\frac{1}{\ve}\delta_{\,ab}D^{\,b},\quad
h_{\,a}
=\frac{1}{\mu}\delta_{\,ab}B^{\,b},
$$
so that Maxwell's equations are closed ones. 

In this contact geometric formulation, Maxwell's equations are realized on the 
Legendre submanifold of the vertical space generated by $\psi^{\,\EM}$.   
Physically this submanifold is the subspace where the energy is properly 
chosen and constitutive relations \fr{constitutive-relation-Maxwell-e-h} 
are satisfied. To describe this, 
one introduces the following adapted functions on the fiber space.
\begin{Def}
\label{definition-adapted-function-D-B}
(Adapted functions for $\bfD$-$\bfB$ oriented formulation): 
$$
\Delta_{\,0}^{\zeta\psi^{\,\EM}}
:=\psi^{\,\EM}-\cE,\qquad 
\Delta_{\,a}^{\,\zeta\psi^{\,\EM}}
:=\frac{\partial\psi^{\,\EM}}{\partial D^{\,a}}-e_{\,a},
\qquad 
\Delta_{\,a+3}^{\,\zeta\psi^{\,\EM}}
:=\frac{\partial\psi^{\,\EM}}{\partial B^{\,a}}-h_{\,a},\qquad 
a\in\{1,2,3\}.
$$
\end{Def}
Associated with this set of functions,   the following are introduced.
\begin{Def}
\label{definition-adapted-mixed-forms-D-B}
(Adapted mixed forms for $\bfD$-$\bfB$ oriented formulation): 
Let $\Delta_{\,0}^{\,\zeta\psi^{\,\EM}}\in\Gamma\Lambda_{\,\mbbH,\mbbV}^{0,0}\cK$, 
$\Delta_{\,\bfD\bfe}^{\,\zeta\psi^{\,\EM}}\in\Gamma\Lambda_{\,\mbbH,\mbbV}^{1,0}\cK$ 
and $\Delta_{\,\bfB\bfh}^{\,\zeta\psi^{\,\EM}}\in\Gamma\Lambda_{\,\mbbH,\mbbV}^{1,0}\cK$ be 
such that  
$$
\Delta_{\,\cE}^{\zeta\psi^{\,\EM}}
:=\Delta_{\,0}^{\zeta\psi^{\,\EM}},\qquad
\bDelta_{\,\bfD\bfe}^{\,\zeta\psi^{\,\EM}}
:=\frac{\delta\wt{\psi}_{\,\cZ_{\,0}}^{\,\EM}}{\delta \bfD}-\bfe,
\qquad 
\bDelta_{\,\bfB\bfh}^{\,\zeta\psi^{\,\EM}}
:=\frac{\delta\wt{\psi}_{\,\cZ_{\,0}}^{\,\EM}}{\delta \bfB}-\bfh.
$$
\end{Def}

With the adapted functions, Maxwell's equations are formulated 
in the following space.
\begin{Def}
\label{phase-space-DB}
(Phase space for the $\bfD$-$\bfB$ formulation of Maxwell's equations): 
Let $\cA_{\,\zeta\psi^{\,\EM}}^{\cC}$ be the 
Legendre submanifold of the vertical space generated by $\psi^{\,\EM}$ 
 as 
$$
\cA_{\,\zeta\psi^{\,\EM}}^{\cC}
=\left\{
\left(\,x^{\,\EM},p^{\,\EM},z^{\,\EM}\,\right)\in\pi^{-1}(\zeta)\,\left|\right.
\Delta_{\,0}^{\,\zeta\psi^{\,\EM}}=\Delta_{\,1}^{\,\zeta\psi^{\,\EM}}=\cdots=\Delta_{\,6}^{\,\zeta\psi^{\,\EM}}=0\,
\right\}.
$$ 
Then 
the sub-bundle $(\cA_{\,\psi^{\,\EM}}^{\,\cK},\pi|_{\,\cA_{\,\psi^{\,\EM}}^{\,\cK}},\cZ)$ with 
$\cA_{\,\psi^{\,\EM}}^{\,\cK}:=\bigcup_{\zeta\in\cZ}\cA_{\,\zeta\psi^{\,\EM}}^{\,\cC}$  
is referred to as the phase 
space for the $\bfD$-$\bfB$ formulation of Maxwell's equations 
( see \fr{Legendre-submanifold-fiber-psi} ).
\end{Def}

This phase space can also be written as the adapted mixed forms as follows.
\begin{Lemma}
$$
\left\{\,
\Delta_{\,0}^{\,\zeta\psi^{\,\EM}}
=\Delta_{\,1}^{\,\zeta\psi^{\,\EM}}
=\cdots
=\Delta_{\,6}^{\,\zeta\psi^{\,\EM}}
=0\,
\right\}
=\left\{\,
\Delta_{\,\cE}^{\,\zeta\psi^{\,\EM}}
=\bDelta_{\,\bfD\bfe}^{\,\zeta\psi^{\,\EM}}
=\bDelta_{\,\bfB\bfh}^{\,\zeta\psi^{\,\EM}}
=0\,
\right\}.
$$
\end{Lemma}
\begin{Proof}
It can be proven with Lemma\,\ref{derivative-Psi-psi-D} and 
\ref{derivative-Psi-psi-B}.
\qed 
\end{Proof}

Then the following is one of the main theorems in this paper.
On the phase space for the $\bfD$-$\bfB$ formulation of Maxwell's equations, 
one has the Maxwell's equations.
\begin{Thm}
\label{Theorem-D-B-Maxwell}
(Maxwell's equation without source in media, induction oriented formulation):   
Choose the contact Hamiltonian functional as 
$$
\wt{h}_{\,\psi^{\,\EM}}
=\int_{\cZ_{\,0}}h_{\,\psi^{\,\EM}}\star 1
=\int_{\cZ_{\,0}}\left[\,
\bDelta_{\,\bfD\bfe}^{\,\zeta\psi^{\,\EM}}\wedge
\bF_{\,\psi^{\,\EM}}^{\,\zeta \bfD\bfe}
+\bDelta_{\,\bfB\bfh}^{\,\zeta\psi^{\,\EM}}\wedge
\bF_{\,\psi^{\,\EM}}^{\,\zeta \bfB\bfh}
+\Gamma_{\,\zeta\psi^{\,\EM}}\left(\Delta_{\,\cE}^{\,\zeta\psi^{\,\EM}}
\right)\,\star1\right],
$$
where $h_{\,\psi^{\,\EM}}\in\Gamma\Lambda_{\,\mbbV}^{\,0}\cK$, 
$\bF_{\,\psi^{\,\EM}}^{\,\zeta \bfD\bfe}, 
\bF_{\,\psi^{\,\EM}}^{\,\zeta \bfB\bfh}\in\Gamma\Lambda_{\,\mbbH,\mbbV}^{2,0}\cK$ 
are  
$$
h_{\,\psi^{\,\EM}}
=\star\left[\,
\bDelta_{\,\bfD\bfe}^{\,\zeta\psi^{\,\EM}}
\wedge\bF_{\,\psi^{\,\EM}}^{\,\zeta \bfD\bfe}\,\right]
+\star\left[\,
\bDelta_{\,\bfB\bfh}^{\,\zeta\psi^{\,\EM}}
\wedge\bF_{\,\psi^{\,\EM}}^{\,\zeta \bfB\bfh}\,\right]
+\Gamma_{\,\zeta\psi^{\,\EM}}\left(\Delta_{\,0}^{\,\zeta\psi^{\,\EM}}
\right),
$$
$$
\bF_{\,\psi^{\,\EM}}^{\,\zeta \bfD\bfe}
:=\dr\,\left(\,\frac{1}{\mu}\star\bfB\,\right),\qquad\mbox{and}\qquad
\bF_{\,\psi^{\,\EM}}^{\,\zeta \bfB\bfh}
:=-\,\dr\,\left(\,\frac{1}{\ve}\star\bfD\,\right),
$$
respectively, and $\Gamma_{\,\zeta\psi^{\,\EM}}$ is such that 
$$
\Gamma_{\,\zeta\psi^{\,\EM}}\left(\,\Delta_{\,\cE}^{\,\zeta\psi^{\,\EM}}\,\right)
=\left\{
\begin{array}{ll}
0&\mbox{for}\quad\Delta_{\,\cE}^{\,\zeta\psi^{\,\EM}}=0\\
\mbox{non-zero}&\mbox{for}\quad\Delta_{\,\cE}^{\,\zeta\psi^{\,\EM}}\neq 0
\end{array}.
\right.
$$  
Then the restricted contact Hamiltonian vertical vector field 
$X_{\,\wt{h}_{\psi^{\,\EM}}}|_{\,\wt{h}_{\psi^{\,\EM}}=0}$ 
gives Maxwell's equations without source and 
the Poynting theorem. 
\end{Thm}
\begin{Proof}
In this proof the relation 
$\cA_{\,\zeta\psi^{\,\EM}}^{\,\cC}=\{\wt{h}_{\,\psi^{\EM}}=0\}$ is used. 

To write the component expression of the contact Hamiltonian vertical 
vector field, one rewrites $h_{\psi^{\,\EM}}$. 
Writing 
\beqa
\bDelta_{\,\bfD\bfe}^{\,\zeta\psi^{\,\EM}}
&=&\left(\,
\bDelta_{\,\bfD\bfe}^{\,\zeta\psi^{\,\EM}}\,\right)_{\,a}\,\bsigma^{\,a},\qquad
\star\bF_{\,\psi^{\,\EM}}^{\,\zeta \bfD\bfe}
=\left(\,
\star\bF_{\,\psi^{\,\EM}}^{\,\zeta \bfD\bfe}\,\right)_{\,a}\,\bsigma^{\,a},
\non\\
\bDelta_{\,\bfB\bfh}^{\,\zeta\psi^{\,\EM}}
&=&\left(\,
\bDelta_{\,\bfB\bfh}^{\,\zeta\psi^{\,\EM}}\,\right)_{\,a}\,\bsigma^{\,a},\qquad
\star\bF_{\,\psi^{\,\EM}}^{\,\zeta \bfB\bfh}
=\left(\,\star\bF_{\,\psi^{\,\EM}}^{\,\zeta \bfB\bfh}\,\right)_{\,a}\,
\bsigma^{\,a},
\non
\eeqa
with \fr{Hodge-2-1}, one has that 
\beqa
h_{\psi^{\,\EM}}
&=&g^{-1}\left(\,\bDelta_{\,\bfD\bfe}^{\,\zeta\psi^{\,\EM}},
\star\bF_{\,\psi^{\,\EM}}^{\,\zeta \bfD\bfe}\,\right)
+g^{-1}\left(\,\bDelta_{\,\bfB\bfh}^{\,\zeta\psi^{\,\EM}},
\star\bF_{\,\psi^{\,\EM}}^{\,\zeta \bfB\bfh}\,\right)
+\Gamma_{\,\zeta\psi^{\,\EM}}\left(\Delta_{\,\cE}^{\,\zeta\psi^{\,\EM}}\right)
\non\\
&=&
\delta^{\,ab}\left(\,\bDelta_{\,\bfD\bfe}^{\,\zeta\psi^{\,\EM}}\,\right)_{\,a}
\left(\,\star\bF_{\,\psi^{\,\EM}}^{\,\zeta \bfD\bfe}\,\right)_{\,b}
+\delta^{\,ab}\left(\,\bDelta_{\,\bfB\bfh}^{\,\zeta\psi^{\,\EM}}\,\right)_{\,a}
\left(\,\star\bF_{\,\psi^{\,\EM}}^{\,\zeta \bfB\bfh}\,\right)_{\,b}
+\Gamma_{\,\zeta\psi^{\,\EM}}\left(\Delta_{\,\cE}^{\,\zeta\psi^{\,\EM}}\right).
\non
\eeqa
In addition, it follows from 
\fr{derivative-Psi-psi-D-equations} and 
\fr{derivative-Psi-psi-B-equations}
that 
$$ 
\left(\,\bDelta_{\,\bfD\bfe}^{\,\zeta\psi^{\,\EM}}\,\right)_{\,a}
=\frac{1}{\ve}\delta_{\,ab}\,D^{\,b}-e_{\,a},\quad\mbox{and}\quad 
\left(\,\bDelta_{\,\bfB\bfh}^{\,\zeta\psi^{\,\EM}}\,\right)_{\,a}
=\frac{1}{\mu}\delta_{\,ab}\,B^{\,b}-h_{\,a}.
$$
Then the component expression of the restricted contact vertical vector field 
is obtained from \fr{coordinate-expression-contact-Hamiltonian-vertical-vector} 
as 
\beqa
\left.\dot{D}^{\,a}\right|_{\cA_{\,\zeta\psi^{\,\EM}}^{\,\cC}}
&=&-\,\left.\frac{\partial h_{\psi^{\,\EM}}}{\partial e_{\,a}}
\right|_{\cA_{\,\zeta\psi^{\,\EM}}^{\,\cC}}
=\delta^{\,ab}
\left.\left(\,\star\bF_{\,\psi^{\,\EM}}^{\,\zeta \bfD\bfe}\,\right)_{\,b}
\right|_{\cA_{\,\zeta\psi^{\,\EM}}^{\,\cC}},
\non\\ 
\left.\dot{B}^{\,a}\right|_{\cA_{\,\zeta\psi^{\,\EM}}^{\,\cC}}
&=&-\,\left.\frac{\partial h_{\psi^{\,\EM}}}{\partial h_{\,a}}
\right|_{\cA_{\,\zeta\psi^{\,\EM}}^{\,\cC}}
=\delta^{\,ab}
\left.\left(\,\star\bF_{\,\psi^{\,\EM}}^{\,\zeta \bfB\bfh}\,\right)_{\,b}
\right|_{\cA_{\,\zeta\psi^{\,\EM}}^{\,\cC}},
\non\\
\left.\dot{e}_{\,a}\right|_{\cA_{\,\zeta\psi^{\,\EM}}^{\,\cC}}
&=&\left.\left(\,\frac{\partial h_{\psi^{\,\EM}}}{\partial D^{\,a}}
+e_{\,a}\frac{\partial h_{\psi^{\,\EM}}}{\partial \cE}\,
\right)\right|_{\cA_{\,\zeta\psi^{\,\EM}}^{\,\cC}}
=\frac{1}{\ve}
\left.\left(\star\bF_{\,\psi^{\,\EM}}^{\,\zeta \bfD\bfe}\right)_{\,a}
\right|_{\cA_{\,\zeta\psi^{\,\EM}}^{\,\cC}},
\non\\
\left.\dot{h}_{\,a}\right|_{\cA_{\,\zeta\psi^{\,\EM}}^{\,\cC}}
&=&\left.\left(\,\frac{\partial h_{\psi^{\,\EM}}}{\partial B^{\,a}}
+h_{\,a}\frac{\partial h_{\psi^{\,\EM}}}{\partial \cE}\,\right)
\right|_{\cA_{\,\zeta\psi^{\,\EM}}^{\,\cC}}
=\frac{1}{\mu}
\left.\left(\star\bF_{\,\psi^{\,\EM}}^{\,\zeta \bfB\bfh}\right)_{\,a}
\right|_{\cA_{\,\zeta\psi^{\,\EM}}^{\,\cC}},
\non\\
\left.\dot{\cE}\right|_{\cA_{\,\zeta\psi^{\,\EM}}^{\,\cC}}
&=&\left.\left(h_{\psi^{\,\EM}}
-e_{\,a}\frac{\partial h_{\psi^{\,\EM}}}{\partial e_{\,a}}
-h_{\,a}\frac{\partial h_{\psi^{\,\EM}}}{\partial h_{\,a}}
\right)\right|_{\cA_{\,\zeta\psi^{\,\EM}}^{\,\cC}}
\non\\
&=&\left.\left[\,
g^{-1}\left(\bfe,\star\bF_{\,\psi^{\,\EM}}^{\,\zeta \bfD\bfe}\right)
+g^{-1}\left(\bfh,\star\bF_{\,\psi^{\,\EM}}^{\,\zeta \bfB\bfh}\right)
\,\right]\right|_{\cA_{\,\zeta\psi^{\,\EM}}^{\,\cC}}.
\non
\eeqa
These are equivalent to write 
$$
\dot{\bfD}
=\dr\bfh,\quad
\dot{\bfB}
=-\,\dr\bfe,\quad
\dot{\bfe}
=\frac{1}{\ve}\star\,\dr\bfh,\quad
\dot{\bfh}
=-\,\frac{1}{\mu}\star\,\dr\bfe,\quad\mbox{on}\quad
\cA_{\,\zeta\psi^{\,\EM}}^{\,\cC},
$$
and 
$$
\dot{\cE}
=\dot{\psi}^{\,\EM}
=g^{-1}(\bfe,\star\dr \bfh)-g^{-1}(\bfh,\star\dr \bfe)
=\star\left(\,\bfe\wedge\dr\bfh-\bfh\wedge\dr\bfe\,\right)
=-\,\star\dr \left(\,\bfe\wedge\bfh\,\right)
\quad\mbox{on}\quad
\cA_{\,\zeta\psi^{\,\EM}}^{\,\cC}.
$$
The last equation above yields the Poynting theorem :  
$$
\frac{\dr}{\dr t}\wt{\psi}_{\,\cZ_{\,0}}^{\,\EM}
\left[\bfD,\bfB\right]
=\int_{\,\cZ_{\,0}}\dot{\psi}^{\,\EM}(\bfD,\bfB)\star1
=-\int_{\,\cZ_{\,0}}\dr\left(\,\bfe\wedge\bfh\,\right)
=-\int_{\,\partial\cZ_{\,0}}\bfe\wedge\bfh,
$$
where $\partial\cZ_{\,0}$ is the boundary of $\cZ$, and 
Stokes' formula has been used for the last equality.

So far the discussion above is carried out on $\cA_{\,\zeta\psi}^{\,\cC}$ and 
that is valid for an open covering $U_{\,i}$ containing $\zeta$. 
Taking into account this, one completes the proof. 
\qed
\end{Proof}
\begin{Remark}
The equation involving the Poyinting $2$-form $\bfe\wedge\bfh$ 
expresses the energy-balance for the system. 
\end{Remark}
\subsection{$\bfe$-$\bfh$ oriented formulation}
In this subsection from the given co-energy functional \fr{Hamiltonian-e-h}, 
Maxwell's equations will be formulated.

Impose $\cE=D^{\,a}e_{\,a}+B^{\,a}h_{\,a}-\varphi^{\,\EM}$ and 
the constitutive relations 
\fr{constitutive-relation-Maxwell-D-B} 
$$
\bfD=\frac{\delta\, \wt{\varphi}_{\,\cZ_{\,0}^{\,\EM}}}{\delta\bfe},\qquad
\bfB=\frac{\delta\, \wt{\varphi}_{\,\cZ_{\,0}^{\,\EM}}}{\delta\bfh},
$$
or equivalently, 
$$
\bfD
=\ve\,\star\bfe,\qquad 
\bfB
=\mu\,\star\bfh,\qquad\mbox{or}\qquad
D^{\,a}
=\ve\,\delta^{\,ab}e_{\,b},\qquad
B^{\,a}
=\mu\,\delta^{\,ab}h_{\,b},
$$
so that Maxwell's equations are closed ones. 

Similar to Definition\,\ref{definition-adapted-function-D-B}, 
one defines the following. 
\begin{Def}
\label{definition-adapted-function-e-h}
(Adapted functions for $\bfe$-$\bfh$ oriented formulation): 
$$
\Delta_{\zeta\varphi^{\,\EM}}^{\,0}
:=D^{\,a}e_{\,a}+B^{\,a}h_{\,a}-\varphi^{\,\EM}-\cE,\qquad 
\Delta_{\,\zeta\varphi^{\,\EM}}^{\,a}
:=D^{\,a}-\frac{\partial\varphi^{\,\EM}}{\partial e_{\,a}},
\qquad 
\Delta_{\,\zeta\varphi^{\,\EM}}^{\,a+3}
:=B^{\,a}-\frac{\partial\varphi_{\,\cZ_{\,0}}^{\,\EM}}{\partial h_{\,a}},\qquad 
a\in\{1,2,3\}.
$$
\end{Def}

Associated with this set of definitions,   the following are introduced.
\begin{Def}
\label{definition-adapted-mixed-forms-e-h}
(Adapted mixed forms for $\bfe$-$\bfh$ oriented formulation): 
Let $\Delta_{\,\zeta\varphi^{\,\EM}}^{\,0}\in\Gamma\Lambda_{\,\mbbH,\mbbV}^{\,0,0}\cK$, 
$\bDelta_{\,\zeta\varphi^{\,\EM}}^{\,\bfD\bfe}\in\Gamma\Lambda_{\,\mbbH,\mbbV}^{\,2,0}\cK$ 
and $\bDelta_{\,\zeta\varphi^{\,\EM}}^{\,\bfB\bfh}\in\Gamma\Lambda_{\,\mbbH,\mbbV}^{\,2,0}\cK$ 
be such that  
$$
\Delta_{\zeta\varphi^{\,\EM}}^{\,\cE}
:=\Delta_{\zeta\varphi^{\,\EM}}^{\,0},\qquad
\bDelta_{\,\zeta\varphi^{\,\EM}}^{\,\bfD\bfe}
:=\bfD-\frac{\delta\wt{\varphi}_{\,\cZ_{\,0}}^{\,\EM}}{\delta \bfe},
\qquad 
\bDelta_{\,\zeta\varphi^{\,\EM}}^{\,\bfB\bfh}
:=\bfB-\frac{\delta\wt{\varphi}_{\,\cZ_{\,0}}^{\,\EM}}{\delta \bfh}.
$$
\end{Def}

With the adapted functions, Maxwell's equations are formulated 
in the following space.
\begin{Def}
\label{phase-space-eh}
(Phase space for the $\bfe$-$\bfh$ formulation of Maxwell's equations): 
Let $\cA_{\,\zeta\varphi^{\,\EM}}^{\,\cC}$ be the 
Legendre submanifold of the vertical space generated by $\varphi^{\,\EM}$ 
as 
$$
\cA_{\,\zeta\varphi^{\,\EM}}^{\cC}
=\left\{
\left(\,x^{\,\EM},p^{\,\EM},z^{\,\EM}\,\right)\in\pi^{-1}(\zeta)\,\left|\right.
\Delta_{\,\zeta\varphi^{\,\EM}}^{\,0}=\Delta_{\,\zeta\varphi^{\,\EM}}^{\,1}=\cdots
=\Delta_{\,\zeta\varphi^{\,\EM}}^{\,6}=0\,
\right\}.
$$ 
Then the sub-bundle 
$(\cA_{\,\varphi^{\,\EM}}^{\,\cK},\pi|_{\,\cA_{\,\varphi^{\,\EM}}^{\,\cK}},\cZ)$ with 
$\cA_{\,\varphi^{\,\EM}}^{\,\cK}:=\bigcup_{\zeta\in\cZ}\cA_{\,\zeta\varphi^{\,\EM}}^{\,\cC}$  
is referred to as the phase 
space for the $\bfe$-$\bfh$ formulation of Maxwell's equations 
( see \fr{Legendre-submanifold-fiber-varphi} ).
\end{Def}

This phase space can also be written as the adapted mixed forms as follows.
\begin{Lemma}
$$
\left\{\,
\Delta_{\,\zeta\varphi^{\,\EM}}^{\,0}
=\Delta_{\,\zeta\varphi^{\,\EM}}^{\,1}=\cdots
=\Delta_{\,\zeta\varphi^{\,\EM}}^{\,6}=0\,
\right\}
=\left\{\,
\Delta_{\,\zeta\varphi^{\,\EM}}^{\,\cE}
=\bDelta_{\,\zeta\varphi^{\,\EM}}^{\,\bfD\bfe}
=\bDelta_{\,\zeta\varphi^{\,\EM}}^{\,\bfB\bfh}=0\,
\right\}.
$$
\end{Lemma}
\begin{Proof}
It can be proven with Lemma\,\ref{derivative-Psi-psi-D} and 
\ref{derivative-Psi-psi-B}.
\qed 
\end{Proof}

There exists a relation between the phase space for the $\bfD$-$\bfB$
formulation of Maxwell's equations and that for $\bfe$-$\bfh$ one.
\begin{Proposition}
(Relation between 
$\cA_{\,\zeta\psi^{\,\EM}}^{\,\cC}$ and $\cA_{\,\zeta\varphi^{\,\EM}}^{\,\cC}$): 
The subspace $\cA_{\,\zeta\psi^{\,\EM}}^{\,\cC}$ in Definition\,\ref{phase-space-DB} 
is diffeomorphic to 
$\cA_{\,\zeta\varphi^{\,\EM}}^{\,\cC}$ in Definition\,\ref{phase-space-eh},   
( see also Remark\,\ref{standard-contact-geometry-Legendre-diffeomorphic} ).
\end{Proposition}
\begin{Proof}
One can prove this by observing that 
$\varphi^{\,\EM}$ is the total Legendre transform of $\psi^{\,\EM}$
due to Lemma\,\ref{Legendre-transform-energy-density-function}.
\qed
\end{Proof}

Then the following is the counterpart of 
Theorem\,\ref{Theorem-D-B-Maxwell}, and 
 one of the main theorems in this paper.  
On the phase space for the $\bfe$-$\bfh$ formulation of Maxwell's equations, 
one has the Maxwell's equations.
\begin{Thm}
(Maxwell's equation without source in media, field intensity oriented formulation):   

Choose the contact Hamiltonian functional as 
$$
\wt{h}_{\,\varphi^{\,\EM}}
=\int_{\cZ_{\,0}}h_{\,\varphi^{\,\EM}}\star 1
=\int_{\cZ_{\,0}}\left[\,
\Delta_{\,\zeta\varphi^{\,\EM}}^{\,\bfD\bfe}\wedge\bF_{\,\zeta \bfD\bfe}^{\,\varphi^{\,\EM}}
+\Delta_{\,\zeta\varphi^{\,\EM}}^{\,\bfB\bfh}\wedge\bF_{\,\zeta \bfB\bfh}^{\,\varphi^{\,\EM}}
+\Gamma^{\,\zeta\varphi^{\,\EM}}\left(\Delta_{\,\zeta\varphi^{\,\EM}}^{\,\cE}
\right)\,\star1\right],
$$
where $h_{\,\varphi^{\,\EM}}\in\Gamma\Lambda_{\,\mbbV}^{0}\cK$, 
$\bF_{\,\zeta \bfD\bfe}^{\,\varphi^{\,\EM}}, 
\bF_{\,\zeta \bfB\bfh}^{\,\varphi^{\,\EM}}\in\Gamma\Lambda_{\,\mbbH,\mbbV}^{1,0}\cK$ 
are  
$$
h_{\,\varphi^{\,\EM}}
=\star\left[\,
\bDelta_{\,\zeta\varphi^{\,\EM}}^{\,\bfD\bfe}
\wedge\bF_{\,\zeta \bfD\bfe}^{\,\varphi^{\,\EM}}\,\right]
+\star\left[\,
\bDelta_{\,\zeta\varphi^{\,\EM}}^{\,\bfB\bfh}
\wedge\bF_{\,\zeta \bfB\bfh}^{\,\varphi^{\,\EM}}\,\right]
+\Gamma^{\,\zeta\varphi^{\,\EM}}\left(\Delta_{\,\zeta\varphi^{\,\EM}}^{\,\cE}
\right),
$$
$$
\bF_{\,\zeta \bfD\bfe}^{\,\varphi^{\,\EM}}
:=\frac{1}{\ve}\star\dr\,\bfh,\qquad\mbox{and}\qquad
\bF_{\,\zeta \bfB\bfh}^{\,\varphi^{\,\EM}}
:=-\,\frac{1}{\mu}\star\dr\bfe,
$$
respectively, and $\Gamma^{\,\zeta\varphi^{\,\EM}}$ is such that 
$$
\Gamma^{\,\zeta\varphi^{\,\EM}}\left(\,\Delta_{\,\zeta\varphi^{\,\EM}}^{\,\cE}\,\right)
=\left\{
\begin{array}{ll}
0&\mbox{for}\quad\Delta_{\,\zeta\varphi^{\,\EM}}^{\,\cE}=0\\
\mbox{non-zero}&\mbox{for}\quad\Delta_{\,\zeta\varphi^{\,\EM}}^{\,\cE}\neq 0
\end{array}
\right..
$$  
Then the restricted contact Hamiltonian vertical vector field 
$X_{\,\wt{h}_{\varphi^{\,\EM}}}|_{\,\wt{h}_{\varphi^{\,\EM}}=0}$ 
gives Maxwell's equations without source and 
the Poynting theorem. 
\end{Thm}
\begin{Proof}
A way to prove this is analogous to the proof of 
Theorem\,\ref{Theorem-D-B-Maxwell}.
\qed
\end{Proof}

\section{Information geometry for Maxwell's equations}
It has been shown in Ref.\,\cite{Goto2015} 
that a contact manifold and a strictly convex function induce  
a dually flat space that is used in information geometry.

Since the energy density function and co-energy density function are  
of strictly convex functions due to Lemma\,\ref{convex-psi-varphi}, 
one can introduce a dually flat space on 
a fiber space of a bundle for the Maxwell fields.  
First,  one introduces a metric tensor field as follows. 
\begin{Def}
(Fiber metric tensor field for the Maxwell fields): 
Let $\psi^{\,\EM}$ be an energy density function defined in 
\fr{definition-energy-density-function}. Then  
the metric tensor field 
$g^{\,\zeta\,\EM}=g_{\,ab}^{\,\zeta\,\EM}\dr_{\,\mbbV}\, x^{\,a}\otimes \dr_{\mbbV}\, x^{\,b}$ 
on $\cA_{\,\zeta\psi^{\,\EM}}^{\,\cC}(\subset\pi^{-1}(\zeta),\zeta\in\cB)$ with 
\beq
g_{\,ab}^{\,\zeta\,\EM}
=\frac{\partial^{\,2}\psi^{\,\EM}}{\partial x^{\,a}\partial x^{\,b}},\quad 
a,b\in\{1,\ldots,6\}
\label{definition-fiber-metric-maxwell-field}
\eeq
and $\{x^{\,a}\}=x^{\,\EM}=\{D^{\,1},D^{\,2},D^{\,3},B^{\,1},B^{\,2},B^{\,3}\}$
is referred to as the fiber metric tensor field of $\cA_{\,\zeta\psi^{\,\EM}}^{\,\cC}$ 
for the Maxwell fields. 
\end{Def}
Noticing Lemma\,\ref{Legendre-transform-energy-density-function}, 
one can show the following. 
\begin{Proposition}
(Components of the contravariant metric tensor field for the Maxwell fields): 
The inverse matrix of $\{g_{\,ab}^{\,\zeta\,\EM}\}$ in
\fr{definition-fiber-metric-maxwell-field} is given as 
$$
g_{\,\zeta\,\EM}^{\,ab}
=\frac{\partial^{\,2}\varphi^{\,\EM}}{\partial p_{\,a}\partial p_{\,b}},\quad 
a,b\in\{1,\ldots,6\},
$$
where $\{p_{\,a}\}=p^{\,\EM}=\{e_{\,1},e_{\,2},e_{\,3},h_{\,1},h_{\,2},h_{\,3}\}$.
\end{Proposition}  
\begin{Proof}
A proof is similar to that found in Ref.\,\cite{AN}.
\qed
\end{Proof}

In the standard information geometry there are two special coordinates, and 
analogous coordinates exist for our formulation of Maxwell's equations.
\begin{Proposition}
(Dual coordinates for the Maxwell fields): 
With  
$x^{\,j}=\partial\,\varphi^{\,\EM}/\partial p_{\,j}$, one has
$$
g^{\,\zeta\,\EM}
\left(\,\frac{\partial}{\partial x^{\,b}},\frac{\partial}{\partial p_{\,a}}\,
\right)
=\delta_{\,b}^{\,a},
$$
where $\{\partial/\partial x^{\,a}\},\{\partial/\partial p_{\,a}\}$ 
are vertical vectors fields.
\end{Proposition}  
\begin{Proof}
It follows from 
$$
\frac{\partial x^{\,j}}{\partial p_{\,a}}
=\frac{\partial^2\,\varphi^{\,\EM}}{\partial p_{\,a}\partial p_{\,j}}
=g_{\,\zeta\,\EM}^{\,aj}  
$$
that 
$$
g^{\,\zeta\,\EM}
\left(\,\frac{\partial}{\partial x^{\,b}},\frac{\partial}{\partial p_{\,a}}\,
\right)
=g_{\,ij}^{\,\zeta\,\EM}\delta_{\,b}^{\,i}\frac{\partial x^{\,j}}{\partial p_{\,a}}
=g_{\,ij}^{\,\zeta\,\EM}\delta_{\,b}^{\,i}\,g_{\,\zeta\,\EM}^{\,aj}
=\delta_{\,b}^{\,a}.
$$
\qed
\end{Proof}

\begin{Remark}
The coordinates $x$ and $p$ satisfying the conditions above 
are referred to as the dual coordinates in the standard information geometry.
\end{Remark}

In the standard information geometry, the dual connections are often discussed. 
These also can appear in the present geometry. 
\begin{Def}
(Dual connections on contact manifolds over a base space): 
Let $\nabla^{\,\zeta}$ be a connection 
on the Riemannian manifold $(\cA_{\,\zeta\psi^{\,\EM}}^{\cC},g^{\,\zeta\,\EM})$, and 
$X_{\,\mbbV},Y_{\,\mbbV},Z_{\,\mbbV}$ vertical vector fields. 
If another connection $\nabla^{\,\zeta\,\prime}$ satisfies 
$$
X_{\,\mbbV}\left[\, g^{\,\zeta\,\EM}(Y_{\,\mbbV},Z_{\,\mbbV})\,\right]
=g^{\,\zeta\,\EM}\left(\nabla_{\,X_{\,\mbbV}}^{\,\zeta}Y_{\,\mbbV},Z_{\,\mbbV}\right)+
g^{\,\zeta\,\EM}\left(Y_{\,\mbbV},\nabla_{\,X_{\,\mbbV}}^{\,\zeta\,\prime}Z_{\,\mbbV}\right), 
$$
then the two connections $\nabla^{\,\zeta}$ and $\nabla^{\,\zeta\,\prime}$
are referred to as dual connections with respect to $g^{\,\zeta\,\EM}$. 
\end{Def}

A realization of connection components of dual connections have been known. 
In our present case of 
$\psi^{\,\EM}$ the following is a trivial identity since $\psi^{\,\EM}$ is a 
quadratic function. 
\begin{Proposition}
(Component expression of dual connections in 
contact manifold over a base space):
Defining 
$$
\Gamma_{\,abc}^{\,\zeta\,(\alpha)}
:=\frac{1-\alpha}{2}
\frac{\partial^{\,3}\,\psi^{\,\EM}}{\partial x^{\,a}\partial x^{\,b}\partial x^{\,c}},
\qquad \alpha\in\mbbR,
$$
one has 
$$
\frac{\partial}{\partial x^{\,a}}g_{\,bc}^{\,\zeta\,\EM}
=\Gamma_{\,abc}^{\,\zeta\,(\alpha)}
+\Gamma_{\,acb}^{\,\zeta\,(-\alpha)},
\quad 
a,b\in\{1,\ldots,6\}.
$$
\end{Proposition}
\begin{Proof}
Substituting \fr{definition-fiber-metric-maxwell-field} 
into the left hand side of the equation above, one completes the proof.
\qed
\end{Proof}
\begin{Remark}
The dual connections $\nabla^{\,\zeta}$ and $\nabla^{\,\zeta\,\prime}$ with respect to
$g^{\,\zeta\,\EM}$ are constructed such that 
$$
\nabla^{\,\zeta}_{\partial/\partial x^{\,a}}\frac{\partial}{\partial x^{\,b}}
=\Gamma_{\,ab}^{\,\zeta(\alpha)\, c}\frac{\partial}{\partial x^{\,c}},\qquad
\nabla^{\,\zeta\,\prime}_{\partial/\partial x^{\,a}}\frac{\partial}{\partial x^{\,b}}
=\Gamma_{\,ab}^{\,\zeta(-\alpha)\, c}\frac{\partial}{\partial x^{\,c}},
$$
where $\Gamma_{\,ab}^{\,\zeta\,(\alpha)\,c}$ and $\Gamma_{\,ab}^{\,\zeta\,(-\alpha)\,c}$ are 
such that 
$$
\Gamma_{\,abc}^{\,\zeta\,(\alpha)}
=g_{\,cj}^{\,\zeta\,\EM}\Gamma_{\,ab}^{\,\zeta\,(\alpha)\,j},\qquad
\mbox{and}\qquad
\Gamma_{\,abc}^{\,\zeta\,(-\alpha)}
=g_{\,cj}^{\,\zeta\,\EM}\Gamma_{\,ab}^{\,\zeta\,(-\alpha)\,j}.
$$
\end{Remark}

With discussions above, one finds the following main theorem in this section.
\begin{Thm}
\label{theorem-Maxwell-dually-flat-space}
(Information geometry for Maxwell's equations):  
Maxwell's equations in media without source induce 
the quadruplet 
$(\cA_{\,\zeta,\psi^{\,\EM}}^{\,\cC}, g^{\,\zeta\,\EM},\nabla^{\,\zeta},\nabla^{\,\zeta\,\prime})$.
\end{Thm}

On any Riemannian manifold $(\cM,g)$ with a connection $\nabla$, 
one can find a dual connection $\nabla^{\,\prime}$. Then 
the quadruplet $(\cM,g,\nabla,\nabla^{\,\prime})$ is referred to as a 
dually flat space\cite{AN}. In accordance with this, one can introduce 
such a space in the present geometry as follows. 
\begin{Def}
(Dually flat space for Maxwell's equations):
The quadruplet introduced in 
Theorem\,\ref{theorem-Maxwell-dually-flat-space}
is referred to as a dually flat space for Maxwell's equations.  
\end{Def}

The canonical divergence plays a role in information geometry, and 
that can be defined in the fiber space as follows.
\begin{Def}
(Canonical divergence on fiber space): 
The function 
$\mbbD^{\,\zeta\,\EM}:\cA_{\,\zeta\psi^{\,\EM}}^{\,\cC}\times\cA_{\,\zeta\psi^{\,\EM}}^{\,\cC}\to\mbbR,(\zeta\in\cZ)$ such that 
$$
\mbbD^{\,\zeta\,\EM}\,(\,\xi\,\|\,\xi'\,)
:=\psi^{\,\EM}(\xi)+\varphi^{\,\EM}(\xi')-x^{\,a}|_{\,\xi}\,p_{\,a}|_{\,\xi'}.
$$
is referred to as canonical divergence for the Maxwell fields.
\end{Def}
The generalized Pythagorean theorem plays a role in the standard information 
geometry, and an analogous theorem exists in the present geometry.
\begin{Thm}
(Generalized Pythagorean theorem for the Maxwell fields): 
Let $\xi^{\,\prime},\xi^{\,\prime\prime}$, and $\xi^{\,\prime\prime\prime}$ be points of  
$\cA_{\,\zeta\psi^{\,\EM}}^{\,\cC}$, $\gamma^{\,\zeta}$ the $\nabla^{\,\zeta}$-geodesic 
connecting $\xi^{\,\prime\prime\prime}$ and $\xi^{\,\prime\prime}$, and 
$\gamma^{\,\zeta\,\prime}$ 
the $\nabla^{\,\zeta\prime}$-geodesic connecting $\xi^{\,\prime\prime}$
and $\xi^{\,\prime}$. 
If at the intersection $\xi^{\,\prime\prime}$ the curves $\gamma^{\,\zeta}$ 
and $\gamma^{\,\zeta\,\prime}$ 
are orthogonal with respect to $g^{\,\zeta\,\EM}$, then   
one has that
$$
\mbbD^{\,\zeta\,\EM}\,(\,\xi^{\,\prime\prime\prime}\,\|\,\xi^{\,\prime}\,)
=\mbbD^{\,\zeta\,\EM}\,(\,\xi^{\,\prime\prime\prime}\,\|\,
\xi^{\,\prime\prime}\,)
+\mbbD^{\,\zeta\,\EM}\,(\,\xi^{\,\prime\prime}\,\|\,\xi^{\,\prime}\,).
$$
\end{Thm}
\begin{Proof}
A proof is similar to that found in Ref.\,\cite{AN}.
\qed
\end{Proof}

\section{Concluding remarks}
This paper offers how Maxwell's equations without source in media 
are formulated with contact geometry.
This formulation is based on the theory of fiber bundles, where  
a fiber space is identified with a contact manifold and a base space 
$3$-dimensional Riemannian manifold expressing physical space. 
The Legendre submanifolds of the contact manifold over the base space 
are equivalent to the spaces where constitutive relations 
 and energy relations hold.  
An important step in this formulation is to recognize that electromagnetic 
energy functional can be seen as an analogue of a convex function 
used in convex analysis. From 
this viewpoint Legendre duality has been focused, and then 
the induction oriented formulation and field intensity oriented one 
have been explicitly shown.  
This viewpoint also has naturally yielded 
information geometry of the Maxwell fields. 

There are numbers of extensions that follow from this work. They are, 
for example,   
to develop a geometric theory of Maxwell's equations 
that can deal with external sources and non-standard constitutive 
relations, and   
to apply some theorems in contact topology to Maxwell's equations. 
We believe that these future works stemmed 
from this work will develop the theory for 
the Maxwell fields and its engineering applications. 
\section*{Acknowledgments }
The author would like to thank 
 Ken Umeno (Kyoto University) 
for supporting my work, also thank 
Yosuke Nakata (Shinshu University), 
Minoru Koga (Nagoya University), and 
Tatsuaki Wada (Ibaraki University) 
for giving critical comments on this paper. 


\end{document}